\documentclass[useAMS,usenatbib]{mn2e}

\usepackage{amsfonts}
\usepackage{amsmath}
\usepackage{graphicx}
\begin{document}
   \title[VVDS: Evolution in the HON over z=0-1.3]{The VIMOS-VLT Deep Survey: \\
    {\LARGE Evolution in the Halo Occupation Number since z $\sim$ 1}\thanks{Based on data
          obtained with the European Southern Observatory Very Large Telescope,
         Paranal, Chile, program 070.A-9007(A), and on data obtained at the
          Canada-France-Hawaii-Telescope, operated by the CNRS of France, CNRC
          in Canada and the University of Hawaii.}}

   \author[U. Abbas et al.]{U. Abbas$^{1,26}$\thanks{E-mail: abbas@oato.inaf.it}, 
	    S. de la Torre$^{1,2,10}$, O. Le F\`evre$^{1}$,
	 L. Guzzo$^{2,3,4}$,
	 C. Marinoni$^{5}$, \newauthor
	 B. Meneux$^{3,6}$,
	 A. Pollo$^{7,8}$,
 	 G. Zamorani$^{9}$, 
         D. Bottini$^{10}$,
	 B. Garilli$^{10}$,
	 V. Le Brun$^{1}$, \newauthor
	 D. Maccagni$^{10}$,
	 R. Scaramella$^{11,18}$,
	 M. Scodeggio$^{10}$,
	 L. Tresse$^{1}$,
	 G. Vettolani$^{11}$, \newauthor
	 A. Zanichelli$^{11}$,
	 C. Adami$^{1}$,
	 S. Arnouts$^{25,1}$,
	 S. Bardelli$^{9}$,
	 M. Bolzonella$^{9}$, 
	 A. Cappi$^{9}$, \newauthor
	 S. Charlot$^{4,15}$,
	 P. Ciliegi$^{9}$,  
	 T. Contini$^{14}$,
	 S. Foucaud$^{24}$,
	 P. Franzetti$^{10}$,
	 I. Gavignaud$^{17}$, \newauthor
	 O. Ilbert$^{23}$,
	 A. Iovino$^{2}$,
	 F. Lamareille$^{9}$,
	 H.J. McCracken$^{15,16}$,
	 B. Marano$^{13}$,  
	 A. Mazure$^{1}$, \newauthor
	 R. Merighi$^{9}$, 
	 S. Paltani$^{20,21}$,
	 R. Pell\`o$^{14}$,
	 L. Pozzetti$^{9}$, 
	 M. Radovich$^{12}$, 
	 D. Vergani$^{10}$, \newauthor
	 E. Zucca$^{9}$, 
	 M. Bondi$^{11}$,
	 A. Bongiorno$^{3}$,
	 J. Brinchmann$^{22}$,
	 O. Cucciati$^{1,2}$,
         L. de Ravel$^{1}$, \newauthor
	 L. Gregorini$^{11}$,
	 E. Perez-Montero$^{14}$,
	 Y. Mellier$^{15,16}$,
	 P. Merluzzi$^{12}$ \\ \\
      (List of affiliations can be found after the References)}

  \date{Accepted 2010 March 30. Received 2010 March 22; in original form 2009 November 04}
 
  \pagerange{\pageref{firstpage}--\pageref{lastpage}} \pubyear{2010}

  \maketitle

\label{firstpage}

  \begin{abstract}
    We model the evolution of the mean galaxy occupation of dark-matter halos 
     over the range $0.1<z<1.3$, using the data from the VIMOS-VLT Deep Survey (VVDS).
    The galaxy projected correlation function $w_p(r_p)$ was computed
     for a set of luminosity-limited subsamples and fits to its shape were obtained 
     using two variants of Halo Occupation Distribution models.  
     These provide us with a set of best-fitting parameters, 
     from which we obtain the
     average mass of a halo and average number of galaxies per halo.
    We find that after accounting for the evolution in luminosity 
      and assuming that we are largely following the same population,
     the underlying dark matter halo %exhibits 
     shows a growth in mass with decreasing redshift as expected 
     in a hierarchical structure formation scenario. 
     Using two different HOD models, 
     we see that the halo mass grows by 90 \% over the
     redshift interval z=[0.5,1.0].     
     This is the first time the evolution in halo %such a growth in halo 
     mass at high redshifts has been obtained from a single data survey and it follows the simple
     form seen in N-body simulations with $M(z) = M_0 e^{-\beta z}$, and 
     $\beta = 1.3 \pm 0.30$.
     This provides evidence for a
     rapid accretion phase of massive halos having a present-day mass
     $M_0 \sim 10^{13.5} h^{-1} M_\odot$, 
     with a $m > 0.1 M_0$ merger event occuring between redshifts of 0.5 and 1.0.
     Futhermore, we find that more luminous galaxies are found to occupy more massive halos 
     irrespectively of the redshift.
     Finally, the average number of galaxies per halo 
       shows little increase from redshift z$\sim$ 1.0 to
     z$\sim$ 0.5, with a sharp increase by a factor $\sim$3 from z$\sim$ 0.5 to z$\sim$ 0.1,
     likely due to the dynamical friction of subhalos within their 
     host halos.
   \end{abstract}

%
%________________________________________________________________

\section{Introduction} \label{intro}

   The correlation function of galaxies is a simple yet powerful 
   tool that allows one to constrain cosmological parameters and 
   models of galaxy formation. Furthermore, with the help of high 
   redshift surveys, the evolution in the clustering of galaxies
   allows for a better discrimination between theoretical 
   models degenerate at the present epoch \citep{pea97}. 
   
   Until recently, the galaxy correlation function had been thought to follow
   a power law %(Totsuji \& Kihara 1969, Peebles 1974, Gott \& Turner 1979)
   \citep{tot69,pee74,got79}.
   Subsequently, a departure from the power law on small scales 
   (of the order 1 to a few Mpc/h) in the galaxy correlation
   function was noticed in pioneering surveys of the eighties, e.g. \citep{guz91}, %Guzzo et al. 1991
   and has now been fully confirmed by many 
   large and deep galaxy surveys.  
   These consist of various 
   surveys at low and intermediate redshifts (z $\le$ 1.5)
   such as the SDSS \citep{con02,zeh04},
  %(Connolly et al. 2002, Zehavi et al. 2004), 
   2dFGRS \citep{mag03}, %(Magliocchetti \& Porciani 2003), 
   VVDS \citep{lef05b, pol06},
   COMBO-17 \citep{phl06}, DEEP2 \citep{coi06}   
   and for lyman break galaxies (LBGs) at high redshifts in the SXDS and GOODS surveys 
   \citep{ouc05, lee06}. %(Ouchi et al. 2005, Lee et al. 2006). 

   Earliest measures showing a non-power law for the galaxy correlation function
   were difficult to interpret as they relied heavily upon
   the angular correlation function, $\omega(\theta)$, at low redshifts \citep{con02}. 
   This meant an integration over a wide range of galaxy
   luminosities and redshifts as well as complicated correlations
   between the statistical errors \citep{zeh04}. This was followed by
   measurements of the projected correlation function, $w_p(r_p)$, 
   and $\omega(\theta)$ in larger and deeper surveys. 
   It has been seen that the deviation from a power law becomes more pronounced for bright
   galaxy samples with $L$ $>$ $L_\star$ \citep{coi06,pol06} and   
   for LBGs at high redshifts \citep{lee06}.  
   Likewise, in SPH simulations %and semi-analytic models 
   the luminous and more strongly clustered galaxies show a similar 
   behaviour where the 'kink' is clearer than in the case of the full galaxy
   sample \citep{wei04}.

   Interestingly enough, the correlation function for dark matter in N-body 
   simulations is well known to be non-adherent to a power law (Jenkins et
   al. 1998, Kauffmann et al. 1999, Cooray \& Sheth 2002 and references therein).
   The natural question that arises is how biased is the galaxy distribution 
   with respect to the underlying matter distribution? 
   The overall shape of the dark matter correlation function is mostly unaffected as
   one goes to higher redshifts as seen in SPH simulations \citep{wei04} and
   N-body simulations \citep{jen98}. This is different to what is seen for high-z
   galaxies \citep{ouc05, lee06, zhe07}, where the
   so-called 'break' is more prominent implying that
   the biasing of the galaxy distribution on large scales, spatial exclusion of dark matter
   halos on small scales, alongwith a host of other complex physical
   processes, such as dynamical friction, feedback from supernovae,
   ram-pressure stripping etc. conspire in a non-trivial way to produce
   differences in the galaxy correlation function at different redshifts. 

   The break in the power law can be physically interpreted in the language of
   the halo model (see Cooray \& Sheth 2002 for a detailed review) as the
   transition between two scales - small scales lying within the halo to those
   larger than the halo.
   It is only natural to use a halo-based prescription where galaxies form  by
   the cooling of gas within dark matter halos \citep{whi78}, which are
   bound, virialized clumps of dark matter
   that are roughly 200 times the background density at that time \citep{gun72}. 
   The galaxies occupy dark matter halos following a HOD (halo
   occupation distribution) model. In turn the HOD fully describes the bias in the
   distribution of galaxies with respect to the underlying dark matter
   distribution \citep{ber02}. 

   The motivation for HOD based models arose when it was noticed that the
   clustering of galaxies could be reproduced by populating halos in
   semi-analytic models with galaxies following a particular probability distribution
   cite{kau99,ben00,ber02}.
 % (Kauffmann et al. 1999, Benson et al. 2000, Berlind \& Weinberg 2002).
   Furthermore, it has been seen that without the help of a proper halo based description
   the strong clustering of red galaxies (at z $\simeq$ 3) can be explained by high and
   unrealistic (anywhere in the range of 70-200 galaxies per halo) 
   occupation numbers to match the observed number density  
   and strong clustering of a small number of high mass halos \citep{zhe04}. %(Zheng 2004).

   Recently, several groups have studied the galaxy correlation in light of
   the HOD models \citep{mag03, van03, ham04, zeh05, ouc05, phl06, con06, zhe07}.
   %(Magliocchetti \& Porciani 2003, van den Bosch et al. 2003, 
   %Hamana et al. 2004, Zehavi et al. 2005, 
   %Ouchi et al. 2005, Phleps et al. 2006, Conroy et al. 2006, Zheng et al. 2007). 
   Most of these works have mainly concentrated on obtaining best-fit HOD
   parameters and consequently the evolution in the HOD and information on the
   underlying dark matter distribution. In some cases, data from different
   surveys having different selections were used to study the evolution \citep{con06, zhe07}
   %(Conroy et al. 2006, Zheng et al. 2007).
   The work in this paper complements and extends these analyses by studying
   the evolution in the HOD over a redshift range z $\approx$ 0.1 - 1.3 for a variety
   of luminosity-limited samples but always for data from the same survey,
   the VIMOS VLT Deep Survey (VVDS).
   We also study two different HOD models in order to obtain a better
   understanding on the degeneracies between the various parameters.
   
   The paper is organized as follows. In section~\ref{data} we briefly describe the
   data used from the VVDS survey. This section is followed by
   Section~\ref{model} giving an outline of
   the theoretical framework. 
   Section~\ref{results} describes the fitting procedure
   alongwith the best-fit parameters obtained for the different
   models. The estimates for the average halo mass and number of
   galaxies per halo (galaxy weighted) for the different luminosity-threshold
   samples are also presented.
   Finally, Section~\ref{disc} wraps up with a discussion and conclusion of
   the results.
   
   Throughout we will assume a flat $\Lambda$CDM model for which
   ($\Omega_0,h,\sigma_8$) = (0.3,0.7,0.9) at $z=0$. Here $\Omega_0$ is the
   density in units of critical density today, $h$ is the Hubble constant
   today in units of 100 km s$^{-1}$ Mpc$^{-1}$, and $\sigma_8$ describes the rms 
   fluctuations of the initial field evolved to the present time using linear
   theory and smoothed with a top-hat filter of radius 8 Mpc/$h$.
   All absolute magnitudes are in the AB system.
   
\section{The VVDS Data and its analysis} \label{data}

\subsection{Description of the data}
   The data used in this analysis comes from the First Epoch VVDS 
   (VIMOS-VLT Deep Survey) 0226-04 ``Deep'' field (hereafter VVDS-Deep). 
   A complete description of the data, survey strategy, data reduction and
   primary goals can be found in Le F\`evre et al. (2005a). Here we will simply give a 
   short description of the data used.

   \begin{table*}
     \caption{Properties of the different subsamples}
%     \caption{Different subsamples for selection 1}             % title of Table
     \label{table1}      % is used to refer this table in the text
     \centering                          % used for centering table
     \begin{tabular}{l l l l l c c l l}      
       \hline\hline                 
       $M_B^{thresh}$ & $M_B^{low}$ & $M_B^{high}$ & z range & $\bar{z}$ & $N_{gal}$ & $\bar{n}$  
       & $M_B^{mean}$ & $M_B^*$ \\
          &      &      &         &           &           & $(10^{-3}h^3/Mpc^3)$  &  &  \\
       \hline  % inserts single horizontal line
       $<-18.0$ & $-17.31$ & $-17.77$ & [0.2-0.6] & 0.44  & 959  & $25.81^{+4.29}_{-4.22}$  & -17.61 &  -20.44  \\
                & $-17.66$ & $-18.00$ & [0.5-0.8] & 0.66  & 1650 & $26.48^{+6.43}_{-6.38}$  & -17.86 &  -20.47 \\
       \hline
       $<-18.5$ & $-17.69$ & $-18.27$ & [0.2-0.7] & 0.53  & 1281 & $18.83^{+3.93}_{-3.74}$  & -18.09 & -20.39 \\    
                & $-18.16$ & $-18.50$ & [0.6-0.9] & 0.76  & 1673 & $20.25^{+6.20}_{-5.90}$  & -18.37 &  -20.80 \\
       \hline
       $<-19.0$ & $-18.02$ & $-18.71$ & [0.2-0.8] & 0.59  & 1410 & $13.79^{+3.53}_{-3.25}$  & -18.50 & -20.37  \\
                & $-18.60$ & $-19.00$ & [0.7-1.05] & 0.88 & 1628 & $15.23^{+5.98}_{-5.43}$  & -18.85 & -20.77\\
       \hline
       $<-19.5$ & $-18.35$ & $-19.16$ & [0.2-0.9] & 0.67  & 1541 & $9.56^{+3.02}_{-2.68}$  & -18.93 & -20.56 \\
                & $-19.04$ & $-19.50$ & [0.8-1.2] & 0.99  & 1526 & $11.05^{+5.53}_{-4.73}$  & -19.32 & -20.76 \\
       \hline
       $<-20.0$ & $-18.67$ & $-19.48$ & [0.2-0.9] & 0.67  & 1143 & $6.77^{+2.39}_{-2.05}$  & -19.25 & -20.57 \\
                & $-19.37$ & $-20.00$ & [0.8-1.35] & 1.05  & 1443 & $7.31^{+4.54}_{-3.63}$  & -19.76 & -20.97 \\
       \hline\hline
     \end{tabular}
   \end{table*}

   \begin{figure}
   \centering
   \includegraphics[width=9cm]{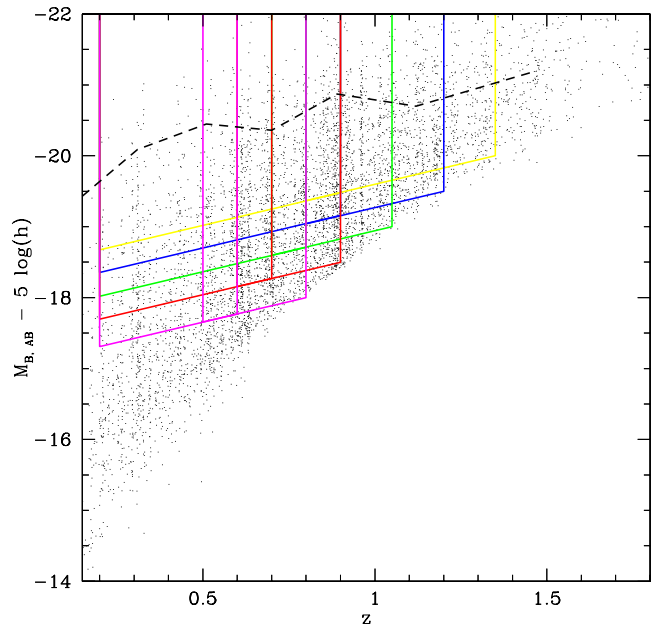}
      \caption{$M_B$ for the VVDS galaxies at different redshifts.
              The solid colored lines denote the various subsamples selected with
              properties mentioned in Table~\ref{table1}. The dashed line
              shows the observed evolution in $M_B^*$.}
         \label{FigMvsZ}
   \end{figure}

   The VVDS-Deep sample is magnitude limited in the $I_{AB}$ band with $17.5 \le I_{AB} \le 
   24$ and covers an area of 0.49 $deg^2$ without any color 
   or shape restrictions imposed. The spectroscopic observations
   were taken with the Visible Multiple-Object Spectrograph (VIMOS, Le F\`evre et al. 2003)
   at the ESO VLT, whereas the Virmos Deep Imaging survey (VDIS) BVRI photometric data 
   \citep{lef04} was obtained with the wide field 12k mosaic camera at the CFHT 
   (Canada-France-Hawaii telescope) and is complete and free
   from surface brightness selection effects \citep{mcc03}.

   The sample contains 6582 galaxies with secure redshifts, i.e. 
   known at a confidence level
   $\ge 80 \%$. These galaxies have a mean redshift of z $\approx$ 0.83.
 %, with an accuracy in the redshift measurements of about $275 km/s$ (Le F\`evre et al. 2004a).
   Fig. \ref{FigMvsZ} shows the absolute magnitude of these galaxies as a function of redshift
   alongwith the different luminosity-threshold subsamples selected, where the
   luminosity threshold is assumed to evolve according to the relation $M_B(z)
   = -1.15z + M_B(z=0)$.  The factor of '-1.15' arises from the redshift evolution of
   the characteristic absolute magnitude,
   $M_B^*$, of galaxies as measured in the luminosity function. 
   This value has been determined using the luminosity function
   measurements obtained within the same sample by Ilbert et al. (2005).

    An evolving luminosity threshold needs to be taken into
    consideration when comparing samples at different epochs, as
    it provides us with statistically similar samples at different redshifts,
    having similar evolved luminosities.   
    Assuming that the global  evolution of galaxies has as a main consequence to 
   increase the global luminosity of galaxies, we follow the evolution of
   galaxies with similar properties on average.
   This falls within the boundaries of standard practice of galaxy evolution studies.

   As our subsamples are nearly volume complete, and as we
   are using all types of galaxies together, we may follow the global increase
   in the halo mass of an average galaxy. However, we do recognize that
   this way of selecting galaxies does not garantee to follow the exact same
   population with cosmic time. Unfortunately, there is no
   single prescription enabling to tag galaxies and exactly follow
   their precursors / descendants. Indeed, if this were possible it would be the solution
   to galaxy evolution.
   To try to quantify the impact of our selection on the average
   halo mass, we have used the Millennium simulation. This will be further discussed in the next
   section.

   Samples using a similar type of selection, i.e. using luminosity thresholds,
   have been extensively studied 
   within a theoretical framework \citep{zeh05, coi06, con06, zhe07}.
   %(Zehavi et al. 2005, Coil et al. 2006, Conroy et al. 2006, Zheng et al. 2007). 
   The corresponding HOD parametrisation 
   requires fewer parameters to be fitted as compared to differential, luminosity binned samples.
   However, this means that one is biased towards increasingly brighter galaxies at higher
   redshifts, simply due to the fact that the sample is selected in apparent magnitude.
%   the fainter galaxies are not as easily detected at high redshift with present instrumentation. 
   As can be seen in Fig. \ref{FigMvsZ}, 
   there is a change of about 5 in the B-band absolute magnitude over the redshift bin 
   z $\approx$ [0.1, 1.5].
   Table~\ref{table1} shows the various properties of the subsamples alongwith their number
   densities.  The galaxy number densities were computed by integrating the luminosity functions,
   derived by Ilbert et al. 2005 on the same galaxy sample and parameterized using Schechter functions.
   The evolution in the best-fit Schechter parameters, $M^*$, $\phi^*$ and $\alpha$ was taken
   into consideration, thereby accounting for luminosity evolution.
   We estimated the errors on the number densities by propagating the errors on the Schechter parameters.
   $M_B^{thresh}$ denotes the evolving absolute magnitude threshold at the
   highest redshift. For each $M_B^{thresh}$ two samples were obtained, one at
   low redshift and another at higher redshift with brighter galaxies
   selected due to the evolving selection cut. The two samples overlap
   slightly in redshift in order to maximise the number of objects. 
   $M_B^{low}$ and $M_B^{high}$ are the absolute magnitudes of the
   evolving cut at the
   lower and higher redshift limits respectively of the redshift range.

 \subsection{Studies with simulations}
    Ideally, we would like to follow statistically the same galaxy population with time
     in order to study the growth of the underlying dark matter halo mass.
   %  In order to study the growth of the underlying dark matter halo mass, one needs to follow 
   %statistically the same galaxy population with time. 
   This is a tricky issue as it is difficult 
   if not impossible to know the exact progenitors of a descendant galaxy population and how to 
   select them. However, by taking care to follow the {\it exact population mix} down to a fixed absolute
   luminosity, we can minimize the bias in the average halo mass due to the presence/lack of faint or 
   bright galaxies. As mentioned above, this is made possible by accurate measurements of 
   galaxy evolution. In order to tackle this issue we use the Milli-Millennium simulation of galaxies 
   having $270^3$ particles in a box of $62.5 Mpc/h$ on its side \citep{spr05}. %(Springel et al. 2005). 
   The simulations retain information on the progenitor
   trees of galaxies making it possible to make a comparison with volume-limited samples.

   Let us select a galaxy sample at high redshift chosen with a luminosity cut-off in the B-band similar
   to what is done in the VVDS data, hereafter called the 'parent sample'.
   This sample is then evolved into two samples at lower redshift, 
   a 'simulated sample' at the lower redshift, having a luminosity cut-off that is 
   evolved and fainter (again similar to what we did in the data), and another sample 
   that contains all the descendant galaxies at the same lower redshift (hereafter the 
   'descendant sample').
   Doing a galaxy to galaxy match between the two samples would tell us how many galaxies 
   in the simulated data set are actual descendants and therefore the same population 
   followed through time and the effects on the underlying average halo mass. 
   Table~\ref{table2} shows the Millennium samples selected having roughly the same mean redshift and mean
   absolute magnitude, $M_B$, as in the VVDS data sample of Table~\ref{table1}. 

   \begin{table*}
     \caption{Simulated data sets vs. descendants}
%     \caption{Different subsamples for selection 1}             % title of Table
     \label{table2}      % is used to refer this table in the text
     \centering                          % used for centering table
     \begin{tabular}{l l l l l c c l l}      
       \hline\hline                 
       Sample & $\bar{z}$ & $M_B^{mean}$ & Avg. Halo mass & Overlap $\%$ \\
              &           &             & $10^{10}M_\odot/h$ &                \\
       \hline  % inserts single horizontal line
        parent     & $0.69$ & $-17.86$  & $127.04$  &   \\
        simulated  & $0.46$ & $-17.61$  & $164.84$  &  \\
        descendant & $0.46$ & $-17.71$  & $176.21$  &  $86.96$   \\
       \hline
        parent     & $0.76$ & $-18.37$ & $135.56$  &     \\
        simulated  & $0.51$ & $-18.09$ & $181.83$  &       \\
        descendant & $0.51$ & $-18.17$ & $199.49$  & $84.03$ &  \\
       \hline
        parent     & $0.90$ & $-18.85$ & $125.82$  &     \\
        simulated  & $0.56$ & $-18.50$ & $192.06$  &     \\
        descendant & $0.56$ & $-18.57$ & $219.35$  & $82.10$ \\
       \hline
        parent     & $0.99$ & $-19.32$ & $125.60$ &      \\
        simulated  & $0.69$ & $-18.93$ & $177.15$ &       \\
        descendant & $0.69$ & $-19.05$ & $202.82$ &  $80.43$   \\
       \hline
        parent     & $1.08$ & $-19.76$ & $126.24$   &       \\
        simulated   & $0.69$ & $-19.25$ & $191.36$ &  \\
        descendant & $0.69$ & $-19.42$ & $232.73$ &  $77.12$ &  \\
       \hline\hline
     \end{tabular}
   \end{table*}
   
   We find that at worst 77\% of the $M<-20$ and at best 87\% of the $M<-18$ 
   simulated luminosity threshold sample are actual descendants at lower redshifts.
   From Table~\ref{table2} we can also study the effect of the selections on the 
   underlying average halo mass. It can be seen that typically the underlying halos
   in the descendant sample are heavier than those in the simulated sample.
   After taking a closer look at the descendant sample we noted that even though
   there are a larger number of fainter galaxies, there are also slightly more bright 
   galaxies that lead to a slightly higher average magnitude.
   The combination of faint satellite galaxies residing in massive halos and fewer 
   galaxies of intermediate luminosity, likely lead to a descendant sample
   with more massive halos and slightly brighter galaxies on average than the simulated sample.
   This possibly causes a lower overlap between the simulated and descendant samples
   for the brightest samples.
   
   The Millennium simulation shows that a growth in halo mass detected in the data would be
   underestimated with respect to what could be seen ideally. 
   The underestimation in mass is of the order of roughly 10 $\%$, and therefore a measure 
   in the growth of mass of a halo can be mainly attributed to the hierarchical formation of 
   structure and not due to the typology of the selection (taking into consideration the high overlap 
   between the simulated and descendant samples).

 \subsection{The correlation function}

   The redshift-space correlation functions for the different luminosity 
   threshold samples have
   been computed via the Landy \& Szalay (1993) estimator:
   \begin{equation}
     \xi(r_p,\pi) = \frac{N_R(N_R - 1)}{N_G(N_G - 1)}\frac{GG(r_p,\pi)}{RR(r_p,\pi)}
                  - \frac{N_R - 1}{N_G}\frac{GR(r_p,\pi)}{RR(r_p,\pi)} + 1
   \end{equation}
   where $N_G$ and $N_R$ are respectively the total number of galaxies and randomly
   distributed points in the same survey. $GG(r_p,\pi)$ is the number of distinct
   galaxy-galaxy pairs with separations lying in the interval ($\pi$,$\pi + d\pi$) in the radial
   direction and ($r_p$,$r_p + d r_p$) perpendicular to the line of sight.
   Likewise, $RR(r_p,\pi)$ and $GR(r_p,\pi)$ are the number of random-random
   pairs and galaxy-random pairs respectively in the same interval. 
   
   In order to avoid redshift space distortions, $\xi(r_p,\pi)$ has been
   integrated along the line of sight to obtain the projected correlation
   function \citep{dav83}: %(Davis \& Peebles 1983):
   \begin{equation}
     \begin{split}
       \omega_p(r_p) &= 2 \int_0^\infty \xi(r_p,\pi) \rm{d}\pi \\
       &= 2 \int_0^\infty \xi\left[\sqrt{r_p^2 + y^2}\right] \rm{d}y,
     \end{split}
     \label{Eq_wprp}
   \end{equation}
   where $\xi(r)$ is the real space correlation function with $r$ =
   $\sqrt{r_p^2 + y^2}$. 
   %The upper limit is chosen to be finite ($r_{max}$=20 Mpc/$h$)
   %in order to avoid noise caused by uncorrelated distant pairs. 
   The measurements using the same sample impose a similar
   upper limit \citep{pol05, pol06, men06}.
   Pollo et al. (2005) found that $\omega_p(r_p)$ is quite insensitive to 
   $\pi_{max}$ in the range of $15 Mpc/h < \pi_{max} < 25 Mpc/h$ for $r_p< 10 Mpc/h$. 
   Too small a value for this limit would cause an underestimation 
   of the small-scale power, and too large a value would introduce noise.
   After several experiments, the optimal value of $\pi_{max}$=20 Mpc/$h$ has been
   adopted.
   
    The errors have been estimated using bootstrap resampling of the data, 
     which consists of computing the variance of $w_p(r_p)$ in $N_{real}$ 
     bootstrap realizations of the sample.
     Each realization is obtained by randomly selecting a subset of 
     galaxies from the data sample 
   allowing for repetitions. A correction factor is then applied to account for the 
   underestimation of the errors obtained using this technique. This correction factor
   has been calibrated on mock samples to match the ensemble error (accounting for cosmic
   variance) of the simulated mock samples (see Pollo et al. 2006).

   \begin{figure*}
   \centering
   $\begin{array}{cc}
     \includegraphics[width=0.52\textwidth]{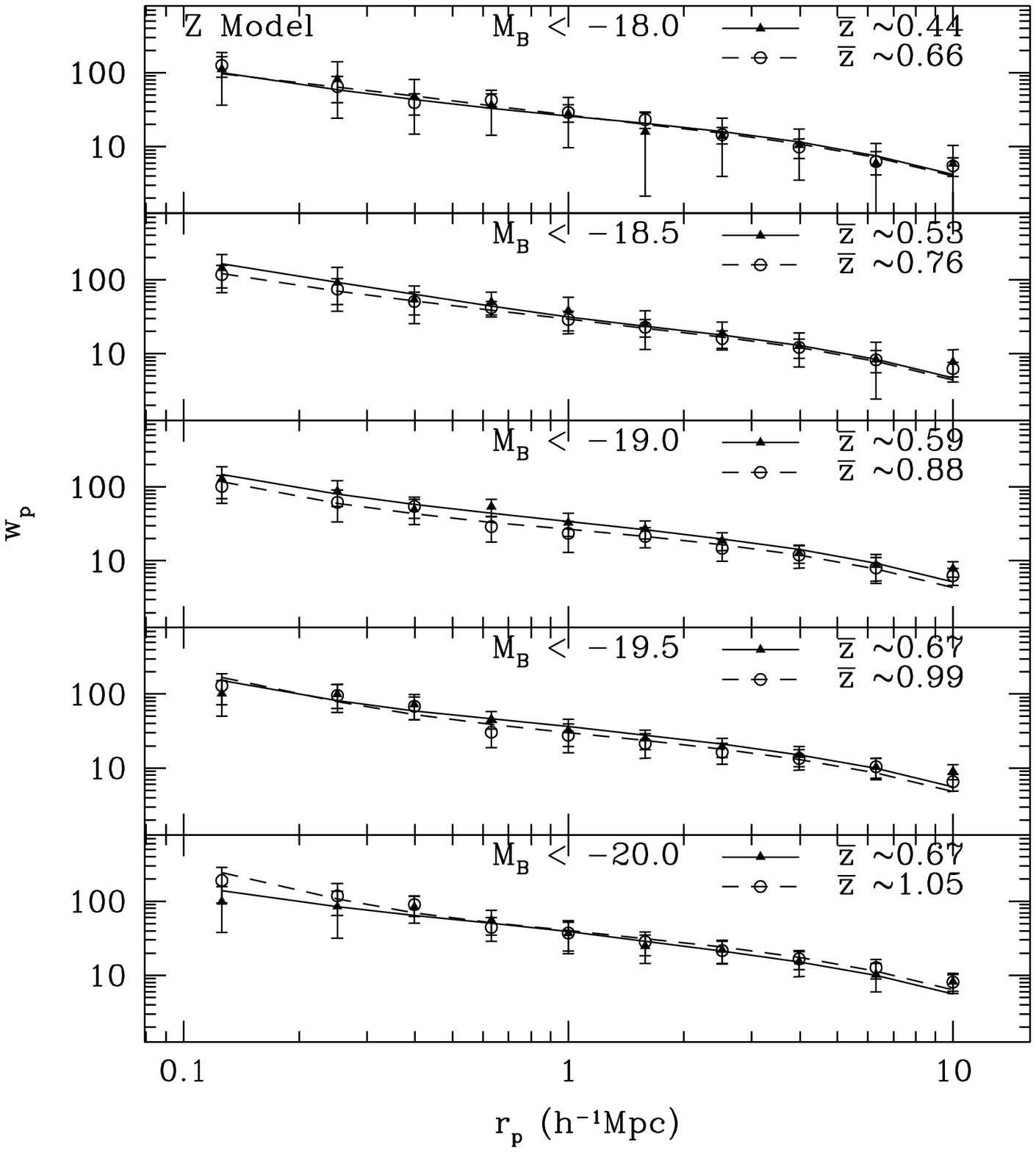} &
     \includegraphics[width=0.52\textwidth]{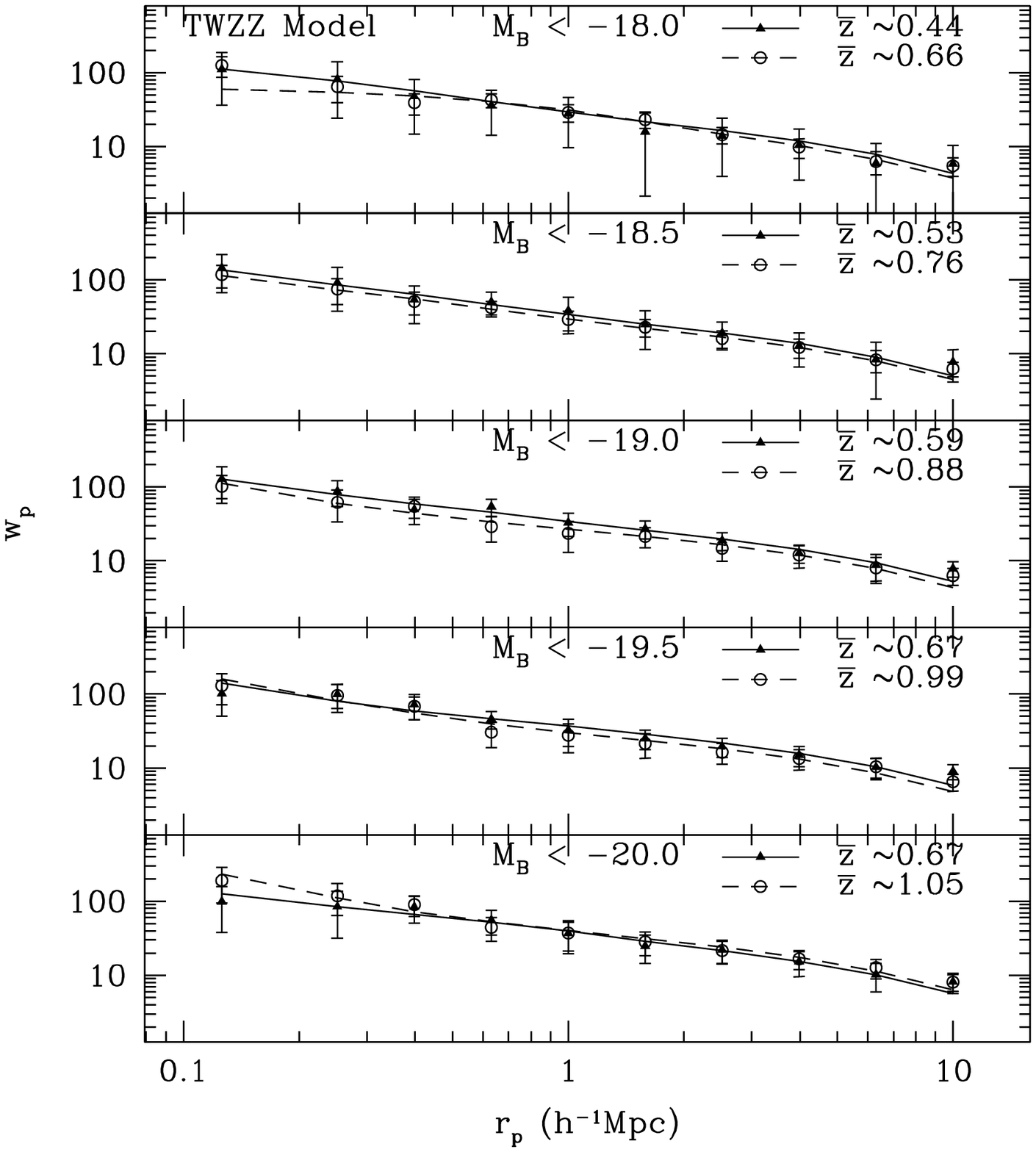} \\
   \end{array}$
   \caption{The correlation function for the various luminosity threshold
            samples. The symbols and error bars denote the measurements
            from the VVDS-Deep. The lines present the best-fit halo model
            for Z model of Zehavi et al. (2005) in the left panel and for 
	    TWZZ model of Tinker et al. (2005) in the right panel.
   }
   \label{Fig_wprp}
   \end{figure*}

\section{Analytical Modeling} \label{model}

\subsection{The Halo model}
   The analytical model is based on the halo model (see Cooray \& Sheth 2002, for a review),
   here we will briefly mention the main ingredients. All mass is assumed to
   be bound up into dark matter halos having a range of masses which in turn
   host galaxies.

   In this model the power spectrum, P(k), and/or correlation function, $\xi(r)$,
   of the galaxies (which are fourier transform pairs) 
   can be written as the sum of two terms. 
   One that dominates on non-linear scales -smaller than the size of a halo, and the other
   term becoming significant on larger linear scales, known as the 1-halo and 2-halo terms
   respectively. The 1 halo term arises from pairs of galaxies lying within the
   same halo, whereas pairs of galaxies lying in different halos contribute
   to the 2 halo term.
   In fourier space this can be written as,

   \begin{equation}
     P(k) = P_{1h}(k) + P_{2h}(k),
   \end{equation}
   where,
   \begin{equation}
     \begin{split}
       P_{1h}^{gal}(k,z) &=  {1 \over\bar n_{gal}^2} \int_0^{\infty} dm\, n(m,z)   
       \langle N_g(N_g-1)|m\rangle\,|u(k|m)|^2, \\
       P_{2h}^{gal}(k,z) &=  \left({1\over\bar n_{gal}}\int_0^{M_{max}} dm\, n(m,z)   
       \langle N_g|m\rangle\,b(m,z)\,|u(k|m)|
         \right)^2 \\
	 &\qquad \qquad \times  P_{lin}(k,z), \\      
     \end{split}
   \end{equation}
   and $\bar n_{gal}$ is the number density of galaxies:  
   \begin{equation}
     \label{eq_ngal}
     \begin{split}
       \bar n_{gal} = \int_0^{\infty}
       \! dm\,n(m,z) \langle N_{g}|m\rangle.
     \end{split}
   \end{equation}
   Here, $\langle N_{g}|m\rangle$ is the 
   average number of galaxies occupying a halo of mass $m$, $u(k|m)$ is the
   halo density profile in fourier space, $n(m,z)$ is the number density of halos of
   mass $m$, $b(m,z)$ is the bias factor which describes the strength of halo clustering, 
   and $P_{lin}(k,z)$ is the power spectrum of the mass in linear theory all
   at a given redshift $z$. 
   The upper limit of integration, $M_{max}$, approximately accounts for the
   halo exclusion effect (different halos cannot overlap) by suppressing the 2-halo term at small scales.
   Following Zehavi et al. (2004), $M_{max}$ is the mass of the halo with
   virial radius $r/2$. 
   One can also calculate the one-halo term for the correlation function 
   exactly in real space, which is the approach we
   have taken. For more details we refer the reader to Berlind \& Weinberg (2002). 
   The two-halo term has been computed in k-space and then fourier transformed
   to obtain the correlation function.
    Then the projected correlation function is obtained as in 
    Equation~\ref{Eq_wprp}. Similarly to the data the upper limit is chosen to be 
    finite ($r_{max}$=20 Mpc/$h$) in order to avoid noise caused by uncorrelated distant pairs.

   We assume that the density profiles of halos have the form described by
   Navarro et al. (1997),  with a halo concentration parameter 
   $c(M) = 11(M/M_*)^{-0.13}$ to account for the definition of halos as spheres enclosing
   200 times the background density \citep{zeh05} and where $\sigma(M_*,z) = 
   \delta_{sc}$ ($\sigma$ and $\delta_{sc}$ are defined below).
   The halo abundances and clustering are
   described by the Sheth \& Tormen (1999) parameterization:
   \begin{equation}
     \begin{split}
       \frac{m}{\bar\rho} n(m,z) {\rm d}m
       &= f(\nu) {\rm d} \nu \\ %f(m) {\rm d}m = f(\nu) {\rm d} \nu \\
       &= \frac{{\rm d}\nu^2}{\nu^2} \sqrt{\frac{a \nu^2}{2 \pi}}
       exp\left(-\frac{a \nu^2}{2}\right) A [ 1+ (a \nu^2)^{-p}] \\
       b(m,z) &= 1 + \frac{a \nu^2 -1}{\delta_{sc}} - \frac{2p/\delta_{sc}}{1 +
       (a \nu^2)^p} \\
       \nu &= \frac{\delta_{sc}}{\sigma(m,z)},
     \end{split}
   \end{equation}
   allowing us to write the background density as,
   \begin{equation}
     \bar\rho = \int {\rm d}m \, n(m,z)\, m.
   \end{equation}
   with $\delta_{sc}$ being the critical density required for spherical collapse,
   extrapolated using linear theory to the present time ($=1.686$,  ignoring 
   the weak cosmological dependence), 
   %for an Einstein de-Sitter model), 
   a $\approx$ 0.71, p $=$ 0.3, and A $\approx$ 0.322. $\sigma(m,z)$ is
   the rms value of the initial fluctuation field when smoothed with a top hat
   filter and extrapolated to the present time using linear theory.
   \begin{equation}
     \sigma(m,z) = \sigma(m,0) \frac{G(z)}{G(0)}
   \end{equation}
   where $G$ is the growth factor from Carroll, Press \& Turner (1992).

\subsection{The HOD models}

   We will consider two similar HOD models. The first one is based on the
   model used in Zehavi et al. (2005) (hereafter Z model) to compare to
   the SDSS data, and is motivated by Kravtsov et al. (2004).
   \begin{equation}
     \begin{split}
       \langle N_g|m \rangle &= 1 + \left(\frac{m}{M_1}\right)^\alpha 
                              \qquad \rm{for \,\, m > M_{min}} \\
                             &= 0   \qquad \qquad \qquad     \rm{otherwise} 
     \end{split}
   \end{equation}
   The second model was proposed by Tinker et al. (2005) (hereafter TWZZ model) 
   and is given by,
   \begin{equation}
     \begin{split}
       \langle N_g|m \rangle &= 1 + \frac{m}{M_1} \rm{exp}\left(-\frac{M_{cut}}{m}
                             \right) \qquad \rm{for\,\, m > M_{min}} \\
                             &= 0    \qquad \qquad \qquad \qquad\qquad   \rm{otherwise} 
     \end{split}
   \end{equation}
   where $M_{min}$ is the minimum mass for a halo to host one central galaxy,
   and $M_1$ is the mass of a halo hosting on average one satellite galaxy.
   The '1' represents one central galaxy placed at the center of mass of
   the parent halo, and the satellite galaxies follow the underlying dark
   matter distribution. TWZZ model has been used to study the HOD for a
   range of redshifts and number densities and found to give results for the
   correlation function in good
   agreement with various redshift surveys in the range $z=[0-5]$ \citep{con06}.
   %(Conroy et al, 2006). 

   Our purpose here will be to obtain the best-fit HOD parameters
   for the two models, and compare the number-weighted halo masses and number of
   galaxies. We have decided to use these models in order to keep things
   simple and easy to interpret. 
    Based on the statistics of the sample and the number of data points used,
    it is best to use HOD models with a minimal number of free parameters adapted to
    the science case at hand, i.e. for this paper, the average halo masses and 
    number galaxies mentioned below.
   The main results that are obtained should
   remain essentially the same irrespective of the HOD model chosen.

    A complementary approach would be to use modeling based on conditional 
    luminosity functions (CLF) or conditional occupation numbers (CON) 
    (e.g. van den Bosch et al. 2003, Yang et al. 2003, Cooray 2006). 
    The present attempt is a first at using a large number of galaxy spectra
    at high redshift to study the evolutionary behaviour of a few properties 
    pertaining to the galaxy and dark matter distribution. These 'few' properties are 
    certainly not comprehensive, and this work can be seen as a starting point for more studies 
    using the sample. Moreover, larger data samples from ongoing and upcoming redshift surveys 
    will certainly provide grounds for extensive studies based on CLF/CON.

\section{Results} \label{results}

\subsection{Results from VVDS}
   The different parameters were allowed to vary within the following ranges: 
   $10 \leq log(M_{min}) \leq 15$, $10 \leq log(M_1) \leq 15$, $10 \leq log(M_{cut}) \leq
   15$, $0.5 \leq \alpha \leq 2.0$. These limits represent reasonable constraints
   on the typical mass of a dark matter halo and the power law slope. The
   minimum mass for a halo to host one central galaxy is usually $\geq 10^{11}
   h^{-1} M_\odot$ for low redshift galaxies \citep{zeh05}, and LBGs at high
   redshifts \citep{ham04}.  
   At the high mass end, the brightest SDSS galaxy samples
   have $M_{min}< 10^{14} h^{-1}M_\odot$. Taking into account our sampling having brightest
   samples of galaxies at high redshifts, in the hierarchical structure formation scenario 
   this then represents an upper limit on the mass. 
   On the other hand, $M_1$ has been found to be $\sim 23 M_{min}$ \citep{zeh05}.  
   Furthermore, power law slopes $\geq$ 1.5 are considered ``artificially
   high'', which generally dominate brighter samples that have fewer satellite
   galaxies on average \citep{con06}.
   The number density obtained using Eq.~\ref{eq_ngal}
   was restricted to lie within 3$\sigma$ from the observed number density
   given in Table~\ref{table1}. 
   The correlation functions for the different luminosity threshold samples
   is shown in Fig.\ref{Fig_wprp} alongwith
   the best fits for the two HOD models obtained with the  MPFIT algorithm \citep{mar09} that
   uses the Levenberg-Marquardt technique \citep{mor78} to solve the non-linear least-squares problem
   using the full covariance matrix.    
   %%Powell's minimization method (Press et al. 1992) using the full covariance matrix. 
   %The fits
   %for the case when $\alpha = 1.1$ (this constraint is explained in the next
   %paragraph) has not been shown as it is very similar
   %to the plot for a flexible $\alpha$.
   %\& \ref{Fig_wprpt}. 
    \begin{table*}
     \caption{Results for Z model}             % with wide limits, no need to show other table
                                               % fits are quite good!
     \label{table3}      % is used to refer this table in the text
     \centering                          
     \begin{tabular}{l c c c c c c l}        
       \hline\hline                
       $M_B^{high}$ & $log_{10}M_{min}$ & $log_{10}M_1$ & $\alpha$ & $log_{10}<M>$ & $<N>$ &
       $\chi^2/dof$ & $\bar{n}_{fit}$ \\    
       \hline                        % inserts single horizontal line
       $<-17.77$  & $11.05 \pm 0.87$ & $12.73 \pm 1.00 $ &
       $0.99 \pm 0.10$ & $13.00 \pm 0.13$ & $1.21 \pm 0.01$ & 1.03 & 37.28 \\
       $<-18.00$   & $11.21 \pm 0.18$ & $13.03 \pm 0.11$  &  
       $1.27 \pm 0.08$ & $12.97 \pm 0.08$ & $1.12 \pm 0.01$ & 1.35 &  26.29 \\
       $<-18.27$  & $11.37 \pm 0.16$ & $12.93 \pm 0.06$ &  
       $1.11 \pm 0.12$ & $13.14 \pm 0.14$ & $1.22 \pm 0.01$ & 0.88 &  20.65 \\
       $<-18.50$  & $11.63 \pm 0.19$ & $13.31 \pm 0.18$ & 
       $1.31 \pm 0.09$ & $12.98 \pm 0.06$ & $1.11 \pm 0.02$ & 0.52 &  11.16 \\
       $<-18.71$  & $11.88 \pm 0.08$ & $13.57 \pm 0.10$ &  
       $1.33 \pm 0.16$ & $13.10 \pm 0.08$ & $1.09 \pm 0.03$ & 1.16 &  6.59 \\
       $<-19.00$   & $11.76 \pm 0.08$ & $13.39 \pm 0.08$ &  
       $1.19 \pm 0.24$ & $12.88 \pm 0.09$ & $1.11 \pm 0.03$ & 0.84 & 8.53 \\
       $<-19.16$  & $12.06 \pm 0.12$ & $13.72 \pm 0.10$ &  
       $1.43 \pm 0.27$ & $13.11 \pm 0.08$ & $1.08 \pm 0.05$ & 1.51 & 4.39 \\
       $ <-19.50$   & $11.92 \pm 0.16$ & $13.43 \pm 0.22$ &  
       $1.22 \pm 0.25$ & $12.94 \pm 0.07$ & $1.12 \pm 0.04$ & 1.24 & 5.99 \\
       $<-19.48$  & $12.06 \pm 0.07$ & $13.78 \pm 0.04$ &  
       $1.66 \pm 0.18$ & $13.13 \pm 0.06$ & $1.06 \pm 0.05$ & 1.08 & 4.36 \\
       $ <-20.00$  & $12.35 \pm 0.11$ & $13.80 \pm 0.16$ &  
       $1.42 \pm 0.41$ & $13.09 \pm 0.04$ & $1.08 \pm 0.10$ & 0.78 & 2.17 \\
       \hline                                 
     \end{tabular}
   \end{table*}

   \begin{table*}
     \caption{Results for TWZZ model}             % title of Table
     \label{table4}      % is used to refer this table in the text
     \centering                          
     \begin{tabular}{l c c c c c c l}        
       \hline\hline              
       $M_B^{high}$ & $log_{10}M_{min}$ & $log_{10}M_1$ & $log_{10}M_{cut}$ & $log(<M>)$ & $<N>$ &
       $\chi^2/dof$ & $\bar{n}_{fit}$ \\    
       \hline                        % inserts single horizontal line
       $<-17.77$ & $11.00 \pm 0.15$ & $12.54 \pm 0.39$ & $13.01 \pm 2.07$ & 
       $13.09 \pm 0.03$ & $1.11 \pm 0.01$ & 1.19 & 38.00 \\   
       $ <-18.00$ & $11.45 \pm 0.19$ & $11.87 \pm 0.71$ & $14.93 \pm 0.62$ & 
       $12.82 \pm 0.05$ & $1.01 \pm 0.02$ & 1.33 & 14.51  \\ %from *.tcfit
       $<-18.27$ & $11.67 \pm 0.13$ & $13.03 \pm 0.22$ & $13.32 \pm 0.75$ & 
       $13.12 \pm 0.01$ & $1.09 \pm 0.02$ & 0.75 & 10.06 \\
       $<-18.50$ & $11.62 \pm 0.08$ & $12.93 \pm 0.08$ & $13.26 \pm 0.24$ & 
       $12.98 \pm 0.01$ & $1.08 \pm 0.02$ & 0.51 & 11.09 \\ 
       $<-18.71$ & $11.88 \pm 0.09$ & $13.22 \pm 0.29$ & $13.45 \pm 0.74$ & 
       $13.10 \pm 0.01$ & $1.07 \pm 0.04$ & 1.16 & 6.40 \\ 
       $ <-19.00$ & $11.74 \pm 0.11$ & $13.22 \pm 0.15$ & $12.62 \pm 0.89$ & 
       $12.88 \pm 0.02$ & $1.09 \pm 0.03$ & 0.81 & 8.71 \\ 
       $<-19.16$ & $12.14 \pm 0.11$ & $13.47 \pm 0.19$ & $13.49 \pm 0.61$ & 
       $13.12 \pm 0.01$ & $1.06 \pm 0.06$ & 1.42 & 3.67 \\ 
       $<-19.50$  & $11.90 \pm 0.16$ & $13.23 \pm 0.35$ & $12.73 \pm 0.82$ & 
       $12.94 \pm 0.02$ & $1.11 \pm 0.04$ & 1.17 & 6.18 \\ 
       $<-19.48$ & $12.07 \pm 0.05$ & $13.15 \pm 0.16$ & $13.97 \pm 0.25$ & 
       $13.13 \pm 0.01$ & $1.04 \pm 0.05$ & 0.97 & 4.18 \\ 
       $<-20.00$ & $12.32 \pm 0.12$ & $13.54 \pm 0.45$ & $13.21 \pm 0.76$ & 
       $13.09 \pm 0.02$ & $1.07 \pm 0.10$ & 0.73 & 2.27 \\ 
       \hline                                 
     \end{tabular}
   \end{table*}

  Tables~\ref{table3} \& \ref{table4} show the best-fit parameters for the 
   two different models obtained by a minimum chi-square estimate,  the value of 
   the reduced $\chi^2$,
   alongwith the average number-weighted halo masses and number of galaxies
   per halo defined as:

   \begin{equation}
     \begin{split}
     \langle M \rangle &= \frac{\int_0^\infty n(m) \langle N_g|m\rangle
       m \, \rm{d}m}{\int_0^\infty n(m) \rm{d}m}, \\
     \langle N \rangle &= \frac{\int_0^\infty n(m) \langle N_g|m\rangle
       \rm{d}m}{\int_0^\infty n(m) \rm{d}m}.
     \end{split}
     \label{Eq_nm}
   \end{equation}

   The generalized chi-square estimate is obtained the usual way adopting: 

   \begin{equation}
     %\chi^2=\sum_{i=0}^{n_{pts}} \frac{[w_p^{obs}(r_{p_i})-w_p^{model}(r_{p_i})]^2}{\sigma^2(r_{p_i})} 
     %+  \frac{[log(n_g^{obs})-log(n_g^{model})]^2}{\sigma^2_{log(n_g)}}
     \chi^2=\sum_{i=0}^{n_{bin}} \sum_{j=0}^{n_{bin}}
              [w_p^{obs}(r_{p_i})-w_p^{model}(r_{p_i})]C_{ij}^{-1}
                [w_p^{obs}(r_{p_j})-w_p^{model}(r_{p_j})] 
   \end{equation}
    where $n_\mathrm{bin}$ is the number of bins and $C_{ij}$ is the 
    covariance of the values of $w_p$ between the $i$th and $j$th bins. 

   The results obtained from both models and given in the tables 
   are found to be in agreement (at least for comparable parameters), which is 
   to be expected as the models are similar. In the case of TWZZ model 
   the power law exponent is kept constant and the number of satellites 
   has a smooth, exponential cut-off. 
   For example, in both cases the value for the
   minimum mass ($M_{min}$) is very similar if not the same. 
   The power law exponent for the Z model several times shows values of $\alpha$ 
    that are quite high ($>1.4$). 
   Artificially high values have been noticed in fits to simulations as well
   \citep{con06} and occur for galaxy samples at high redshifts.
   The 1-$\sigma$ error bars on the parameters are obtained with the MPFIT algorithm.

   \begin{figure*}
     \centering
     $\begin{array}{cc}
     \includegraphics[width=0.5\textwidth]{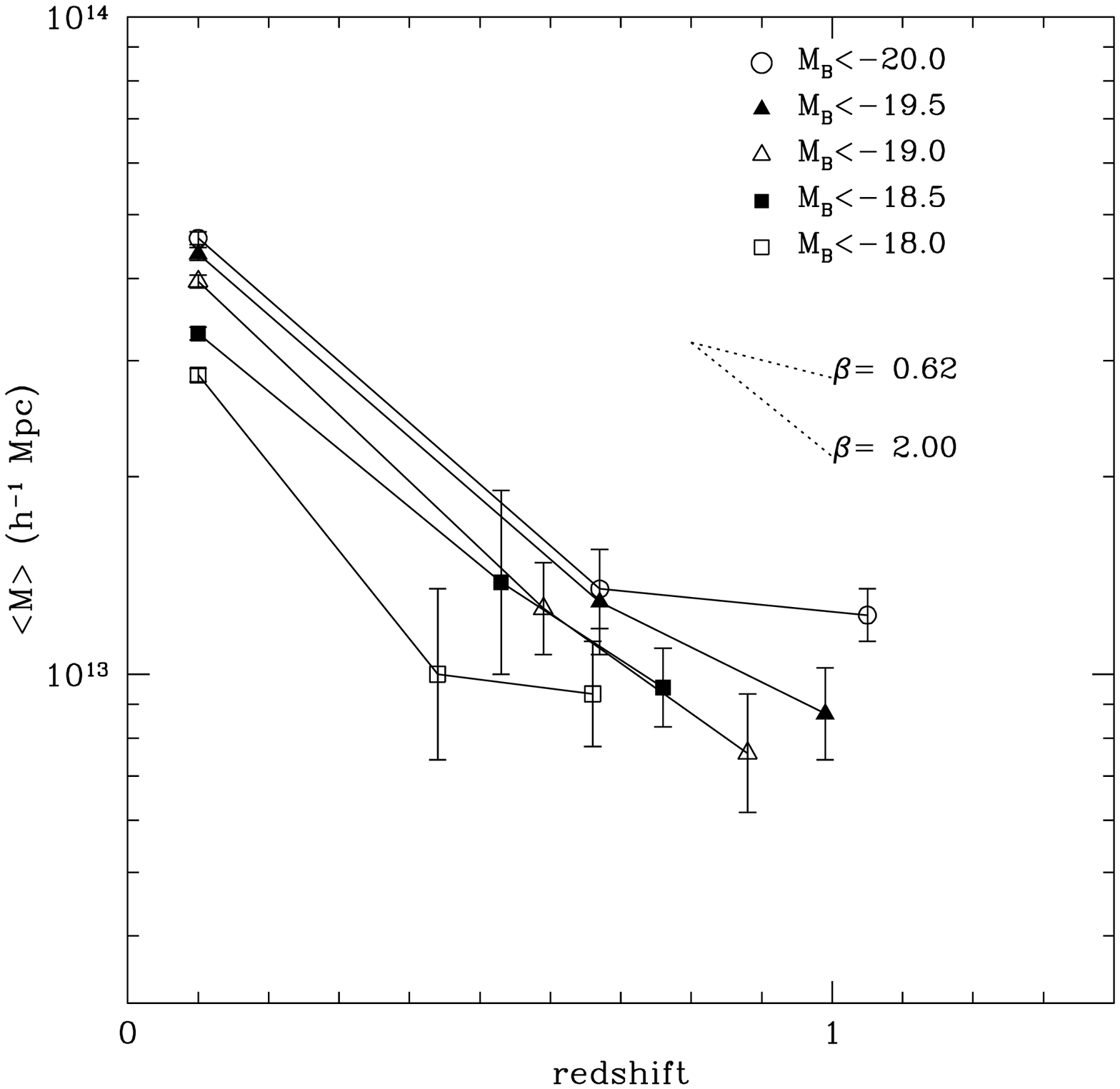} &
     \includegraphics[width=0.5\textwidth]{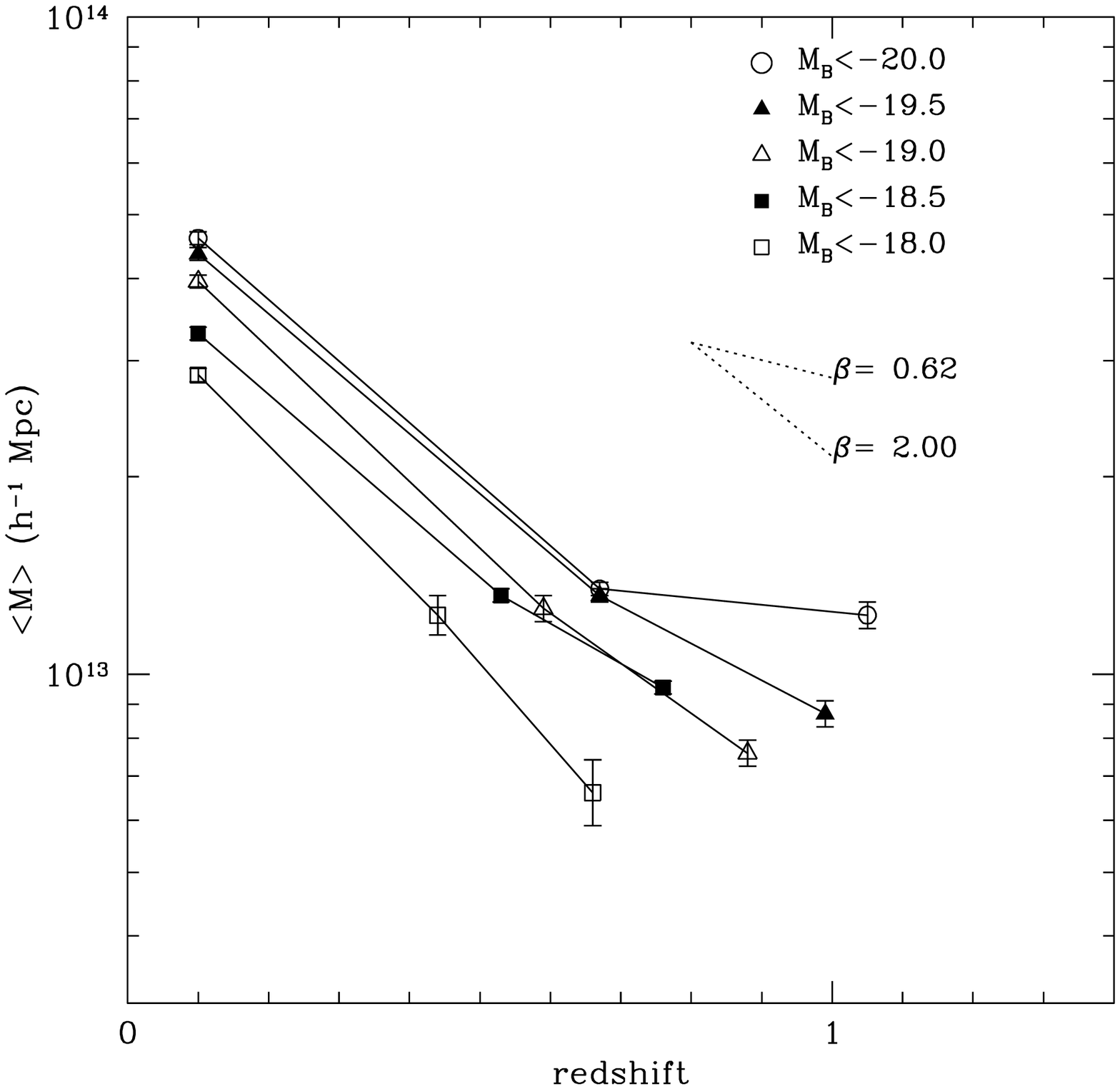} \\
     \end{array}$
     \caption{The evolution in the number-weighted average halo mass 
       given by Eq.\ref{Eq_nm} for various luminosity threshold samples is shown 
%     the difference between the mean absolute magnitude of the sample and the 
%     characteristic magnitude $M_B^*$ 
      for the Z model (left panel) and TWZZ model (right panel). 
      The symbols with error bars are obtained from the VVDS, whereas the low
      redshift symbols at $z=0.1$ with small error bars 
      are from the SDSS. The mass growth can be characterized
      by $M(z) = M_0 e^{-\beta z}$, with $\beta = 1.07 \pm 0.57$ (for the VVDS points),
      $\beta = 1.94 \pm 0.10$ (VVDS + SDSS points) in the case of the Z model, and 
      $\beta = 1.54 \pm 0.13$ (for the VVDS points), $\beta = 2.09 \pm 0.04$ (VVDS + SDSS points)
      for the TWZZ model.
      %for the four brightest samples. 
      The values of $\beta$ =0.62 and 2.00  
      respectively represent the prediction from N-body simulations and the
      VVDS + SDSS samples. 
      %The faintest sample with $M_B < -19.5$   has a value for $\beta > 3$.  
     }
   \label{Fig_Mvsz}
   \end{figure*}

   \begin{figure*}
     \centering
     $\begin{array}{cc}
       \includegraphics[width=0.5\textwidth]{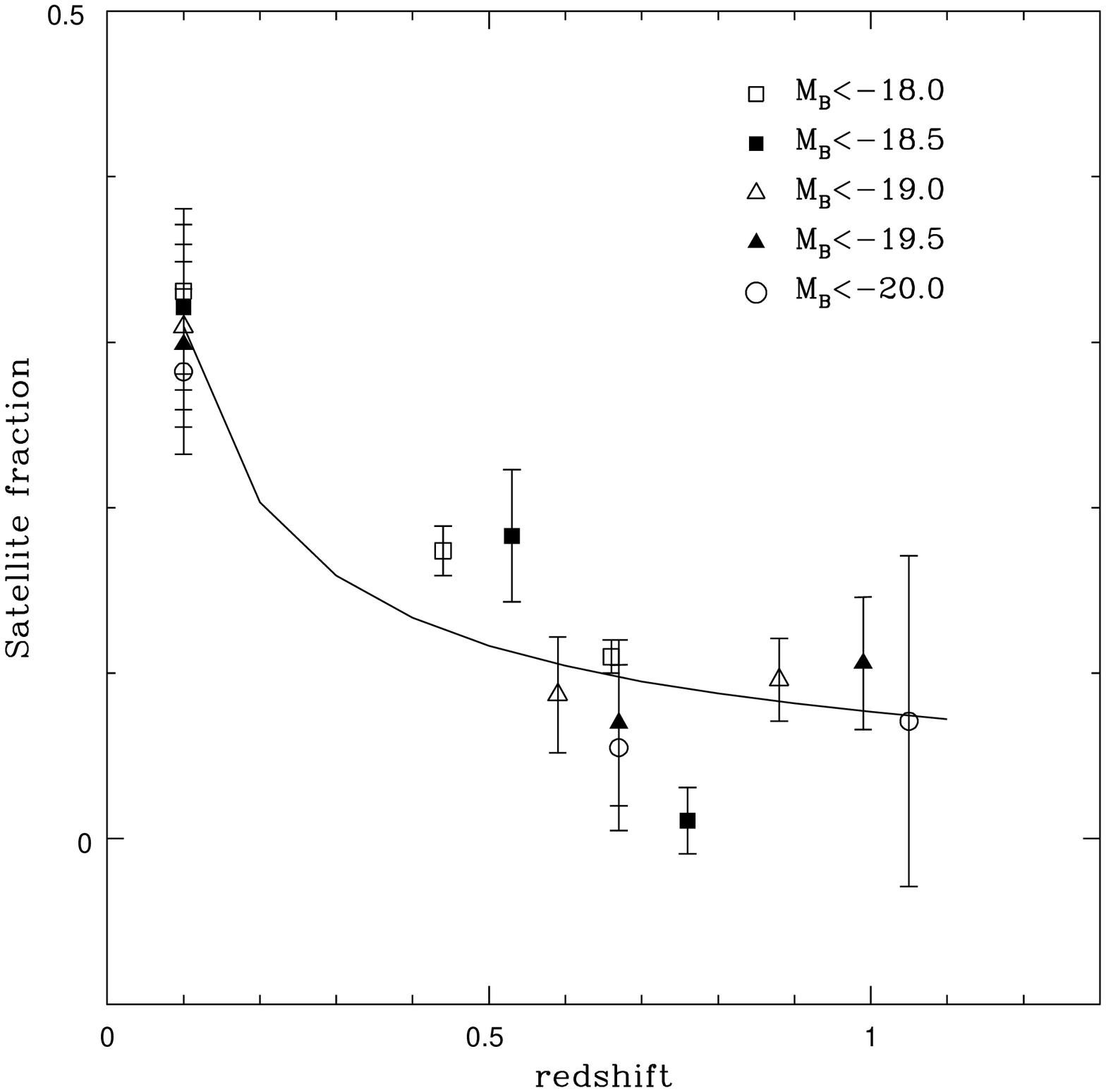} &
       \includegraphics[width=0.5\textwidth]{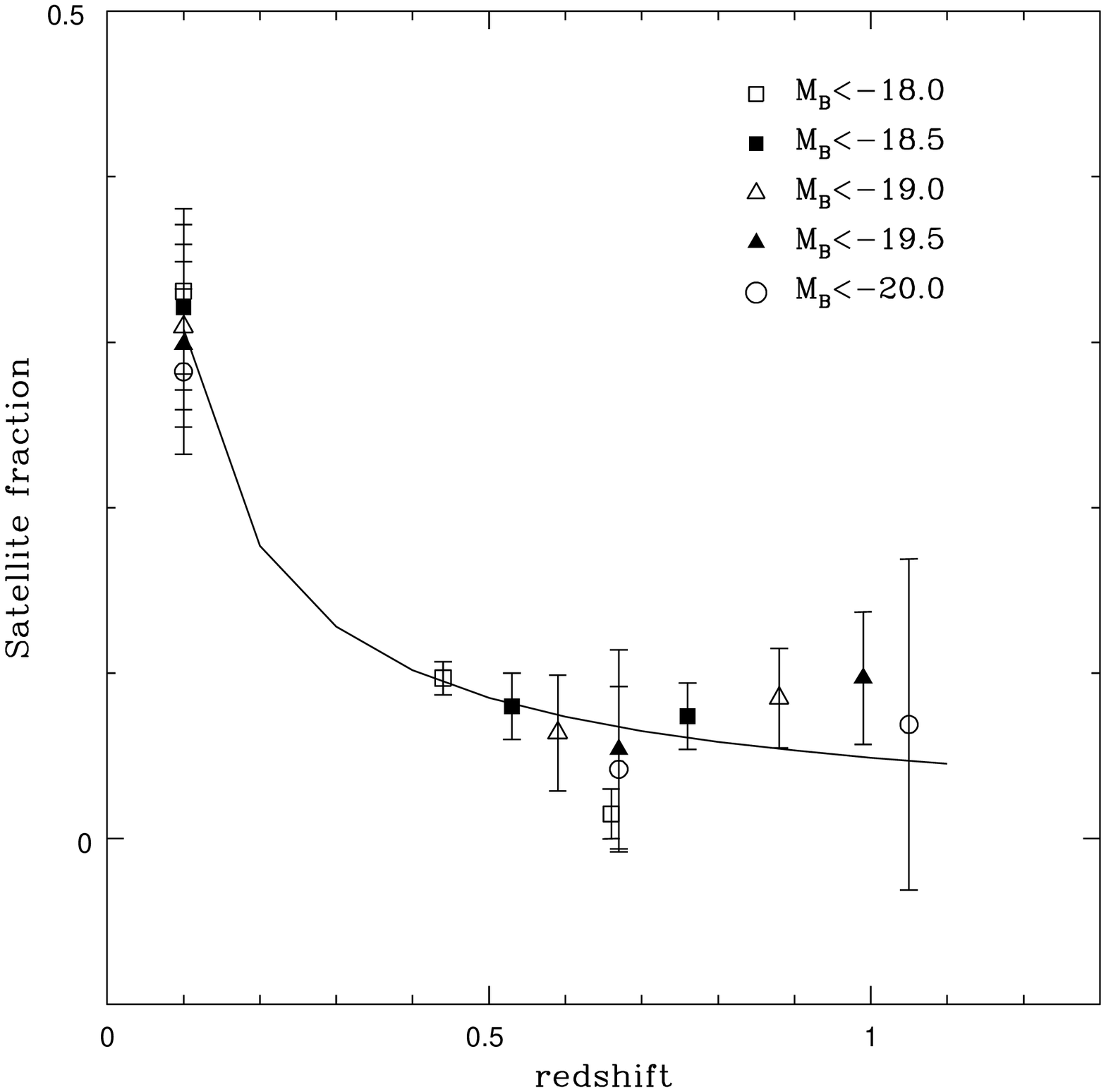} \\
     \end{array}$
     \caption{The evolution in the satellite fraction 
       for various luminosity threshold samples is shown  
%       versus the difference between the mean absolute magnitude of the sample and the 
%       characteristic magnitude $M_B^*$ 
       for the Z model (left panel) and TWZZ model (right panel).
           The symbols with error bars are obtained from the VVDS, whereas the low
      redshift symbols at $z=0.1$ are from the SDSS.  Simply for illustrative purposes, 
      the curves show the reciprocal 
      standard power law behaviour, $y=1/(ax^b)$, of the data, 
      with $a=13.05$ and $b=0.61$ for the Z model, and $a=20.41$ and $b=0.79$ for the TWZZ model.}
   \label{Fig_Nvsz}
   \end{figure*}

   Figure \ref{Fig_Mvsz} shows the number-weighted average halo mass, $<M>$,
   versus the redshift. The symbols with error bars represent the various
   sub-samples selected from the VVDS.
   The error bars are obtained based on error propagation formulas.
   The point at the lowest redshift ($z \sim 0.1$)
   is obtained from the SDSS using the best-fit HOD parameters from Zehavi et
   al. (2005). The mass in this case was calculated using the Z model for the
   luminosity threshold sample having the same $M - M^*$ difference as the samples at
   higher redshift in the VVDS (where the difference in the r-band for the
   SDSS has been converted to the B-band, Ilbert et al. 2005).  
   It can be seen that the halo mass evolves and increases as one goes to lower
   redshifts.  This is an indication of the halo mass growth 
   due to the hierarchical aggregation of matter.  
   We find that on average $<M>$ increases by $~90\%$
   %by a factor $\sim$ 3 
   from redshift $\sim 1$ to $\sim 0.5$, showing that massive halos have a
   rapid accretion phase quite late on, similar to what is expected from N-body
   simulations (Wechsler et al. 2002). 
   As shown in Wechsler et al. (2002) the mass growth can be easily
   characterized by the form $M(z) = M_0 e^{-\beta z}$. 
   The interesting comparison with the addition of low redshift SDSS points
   gives a linear
   minimum $\chi^2$ fit of $\beta \sim 1.94 \pm 0.10$ for
   the Z Model and $\beta \sim 2.09 \pm 0.04$ for the TWZZ model.
   This is to be compared to the 
   predictions of the mass accretion history of halos in N-body simulations and 
   halos generated through PINOCCHIO \citep{mon02, wec02, li07},
   %(Monaco et al. 2002, Wechsler et al. 2002, Li et al. 2007),
   where $\beta \sim 0.62$. One can argue that the direct comparison 
   between data obtained from different rest-frame bands can be tricky 
   and could in part lead to a slight boost in $\beta$, even though 
   %However it should be possible to compare the underlying dark matter halo mass, once
   necessary care has been taken in converting to a common rest-frame band.
   The latter is reflected in the value obtained for $\beta$, using 
   only the VVDS points leads to a smaller value ($\sim 1.54 (1.07)\pm 0.13 (0.57)$ for 
    TWZZ (Z) model) albeit with  
   larger error than that obtained from the extrapolation to smaller redshifts, but slightly more
   consistent with the results from simulations.  

   For samples at similar redshifts we can see that the number-weighted average 
   halo mass increases with the luminosity threshold of the sample reinforcing
   the notion that luminous galaxies occupy massive halos. This is in
   agreement with results obtained from simulations \citep{con06} and for
   LBGs \citep{ouc05, lee06}. %(Ouchi et al. 2005, Lee et al. 2005).

   Figure \ref{Fig_Nvsz} presents the evolution of the galaxy satellite
   fraction, or the average number of satellite galaxies.
    The illustrative reciprocal power law behaviour of the dataset shows  
     relatively little change in the satellite fraction
   (always close to $\sim0.1$ within 1 $\sigma$) over the redshift range of z=[0.5-1.0]. Over z=[0.1-0.5] there is a sharper
    increase by a factor $\sim$3 to the local SDSS value of $\sim$0.3. The evolution
    is mainly accentuated by the SDSS points, although the two lowest redshift VVDS 
    points for the case of the Z model do hint towards an increase with lower redshifts.
   It is possible that the
   sharper upturn is once again caused by the complicated comparison between two
   different data surveys.
   However, here again care has been taken to convert to the appropriate rest-frame band
   when making these comparisons and should not affect the overall trend. 
   The increase in the 
   satellite fraction as one goes to lower redshifts can be explained by 
   the dynamical friction of subhalos within their host halos \citep{con06}. %(Conroy et al. 2006).
   Subhalos are more likely to remain intact within massive halos, whereas in less 
   massive halos they are subject to more dynamical friction and can easily be
   destroyed. The dynamical friction becomes more/less efficient as a function of
   the relative masses of subhalos to distinct halos.
   This is to be compared to recent results obtained by Zheng et al. (2007) who find
   that the evolution of the satellite fraction follows a trend similar to what is seen here. 

   The following Fig. \ref{Fig_hod} shows the evolution in
   the halo occupation, $N_g(m)$, for the extreme luminosity threshold
   samples obtained from the best fit parameters for the two
   models. Evidently, the minimum mass, $M_{min}$ increases with the luminosity of the
   sample as is found locally in the SDSS \citep{zeh05}, again 
   demonstrating that luminous galaxies occupy more massive halos. 
  
   \begin{figure*}
     \centering
     $\begin{array}{cc}
       \includegraphics[width=0.5\textwidth]{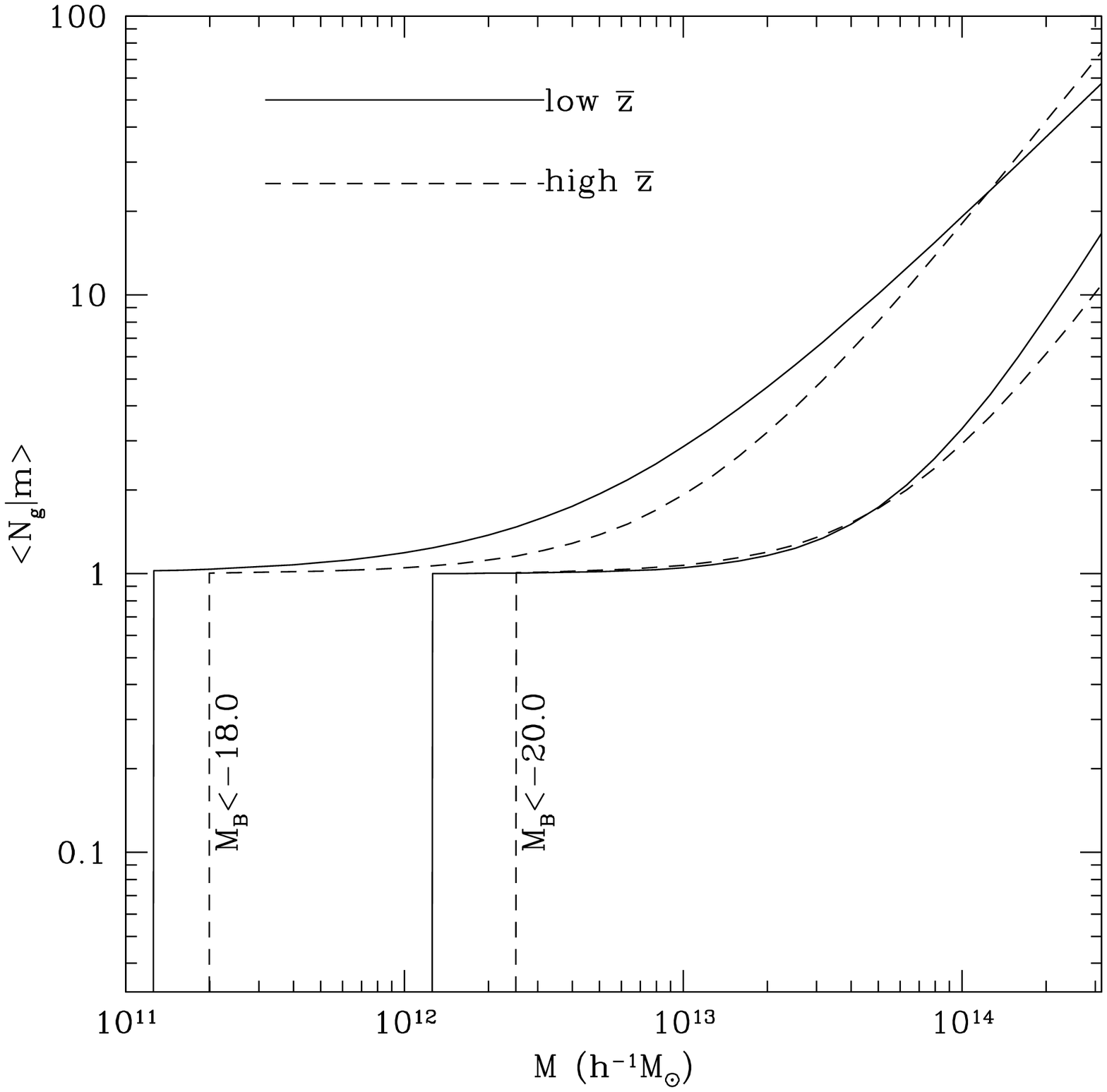} &
       \includegraphics[width=0.5\textwidth]{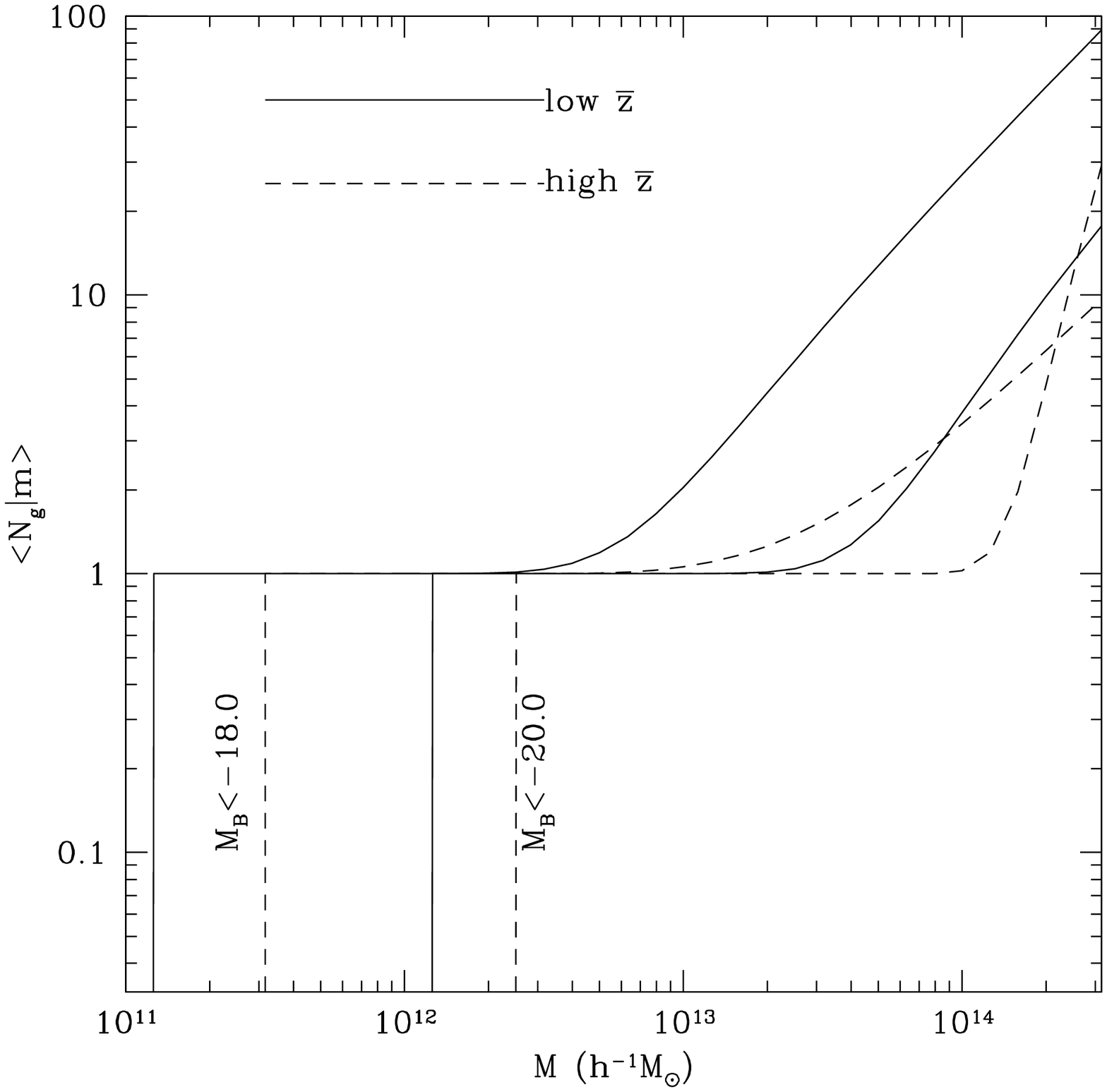} \\       
     \end{array}$ 
     \caption{The evolution of the halo occupation for the extreme
       luminosity threshold samples having $M_B <$ -18, -20. 
       The solid line corresponds to the
       lower redshift sample in each case, and the dashed line to the brighter
       sample at higher redshift both having the same evolved luminosity
       threshold (see Table~\ref{table1}). The solid lines correspond to
       $\bar{z}= 0.44, 0.67$ and the dashed lines to
       $\bar{z}= 0.66, 1.05$ respectively for
       the samples with $M_B <$ -18, -20.
       The left panel depicts Z model and the right panel TWZZ model.}
   \label{Fig_hod}
   \end{figure*}

\subsection{Comparison to SDSS}

\begin{figure*}
     \centering
     \includegraphics[width=15cm]{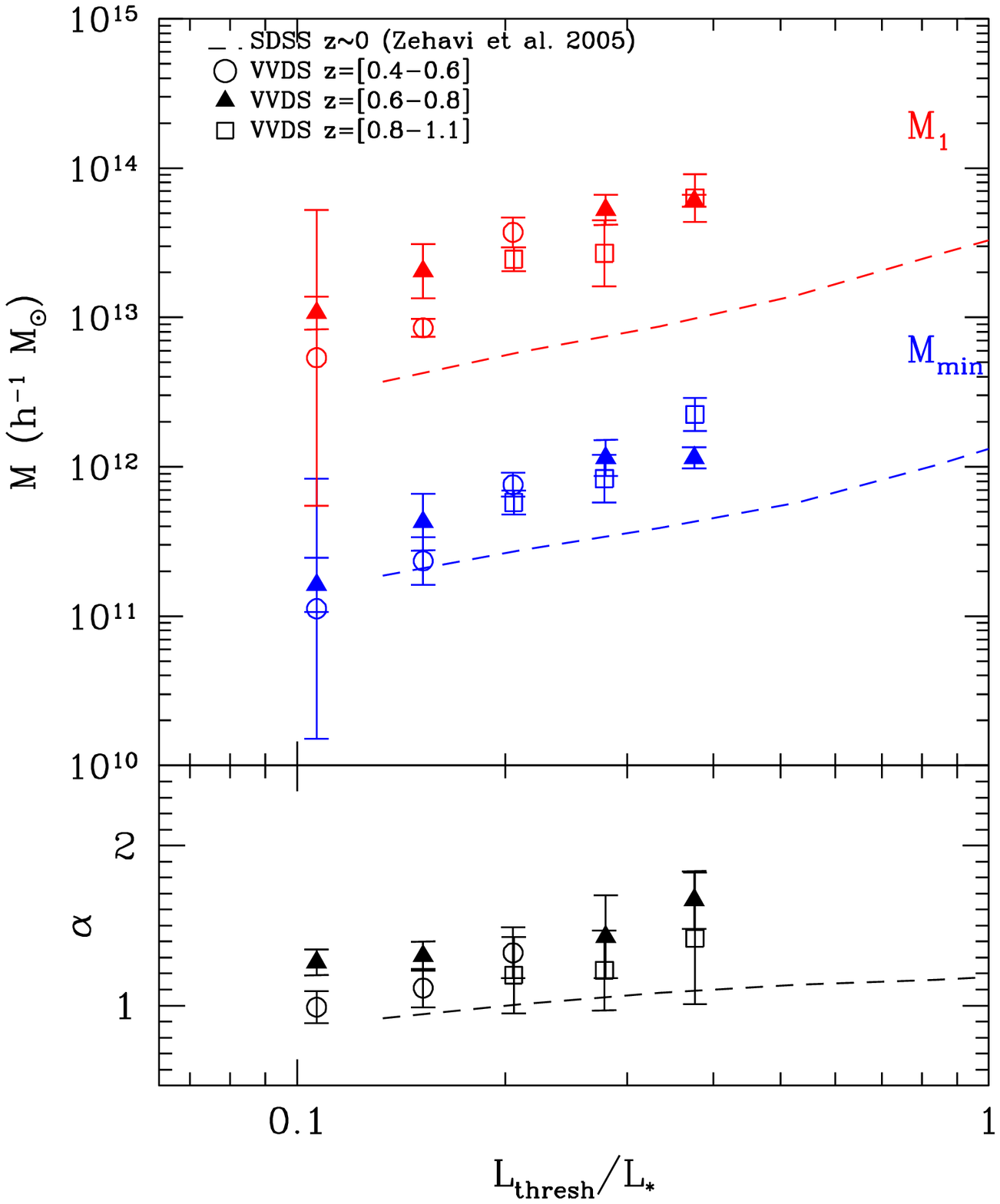}
     \caption{The best fit masses ($M_{min}, M_1$) and power law slope $\alpha$
       of halos for Z model versus $(L_{thresh}/L_*)$. 
       The different symbols 
       represent results for the VVDS, where each set of symbols correspond to 
       samples lying in different redshift ranges (circles, solid triangles,
       and empty squares represent samples having $z = [0.4-0.6], [0.6-0.8], [0.8-1.1]$
       respectively).  
       The superimposed dashed lines depict the SDSS.}
   \label{Fig_comparison}
   \end{figure*}

   In this section we will compare to results for the same HOD model (Z model)
   as used in 
   Zehavi et al. (2005). Figure~\ref{Fig_comparison} shows the comparison
   between the masses of halos that have at least one central galaxy ($M_{min}$) and one
   satellite galaxy ($M_1$) on average as a function of $L_{thresh}/L_*$,
   where the ratio is in the B-band, 
   and $L_{thresh}$ is the luminosity threshold given in Table~\ref{table1}
   ($L_{thresh}$ and $L_*$ are at similar redshifts). Here we will try to compare 
   results obtained at different redshifts.
    For $40 \%$ of the VVDS samples, $M_{min}$ is similar to
   the local SDSS results within the error bars, with the rest of the VVDS samples
   having higher values of $M_{min}$. 
   Generally, the VVDS 
   samples exhibit more massive halo masses, $M_1$, required to host satellite galaxies
   than what is seen locally.
   The value for the power law slope, $\alpha$, is mostly similar to
   that for local galaxies, with the bright intermediate redshift galaxies showing
   a higher slope. 
   We can see that generally the samples with higher values for $\alpha$, 
   also have higher values of $M_1$ and $M_{min}$ than present-day galaxies. 
   
   It is interesting to note the $M_1/M_{min}$ ratio, which is
   on average $\sim$ 45, 
   rather high as compared to the value of $\sim$ 23 for the SDSS galaxies.
   A direct comparison and interpretation of these results is complicated 
   as one is looking at two different surveys taken in different restframe bands. 
   However, we can
   speculate that the high value of the ratio implies that the halo with one
   central galaxy needs to accrete roughly 45 times its mass in order to host
   a satellite galaxy. In other words, a halo of a given mass is likely to have
   fewer satellite galaxies at higher redshifts as opposed to a halo of the same mass
   observed locally.

\section{Discussion \& Conclusions} \label{disc}
%  We have presented results for HOD fits to data from the VVDS survey.

  The comparison of analytical models and data provides useful information
  of how the distribution of galaxies depends on the underlying dark matter.
  Subsequently, the best-fitting parameters obtained as a result of this
  comparison provide physical information regarding the dark matter halos and 
  galaxies. 

   The size of the VVDS dataset allows one to study,
   with a unique sample, the global
  change in the underlying halo properties of an average galaxy down to $z \sim 1$. 
   We attempt to follow the evolution in some properties
  of a magnitude selected sample, evolving the magnitude cut-off based on
  accurate measurements of galaxy evolution.
  We have presented results of the fitting of analytical halo models,  incorporating
  simple HOD models with minimal number of free parameters, 
  to data (in this case the projected 2-point correlation function) from the VVDS survey. 
  This allowed us to study the evolution of the average number weighted halo
  mass and satellite fraction.

  On different scales there are contributions from central-satellite,
  satellite-satellite, and central-central pairs of galaxies to the
  correlation function thereby providing constraints on the evolution of the galaxy satellite fraction.
  The evolution was obtained from data observed in 
  the same restframe band, and provides for simpler interpretations as
  compared to previous studies using data from
  different restframe bands.
  Various luminosity threshold samples at 
  different redshifts were selected and the corresponding best-fit HOD parameters
  for two similar HOD models obtained.
  This is done in order to single out possible degeneracies and inconsistencies with the
  fitting procedure at high redshifts. 
  On the whole, both models are 
  in agreement with each other and show similar trends in evolution.
  The impact of our selection on the average halo mass is addressed using 
  the Millennium simulation. 
  We find that a growth in halo mass as seen in the data 
  could rather be an underestimation of $\sim 10\%$ to what is seen in an 'ideal' sample
  containing all the descendants. Therefore a measure in the growth of mass of a halo
  can be mainly attributed to the hierarchical formation of structure and not 
  due to the typology of the selection.
  
   We find that the number-weighted average halo mass grows by $\sim 90 \%$ 
  from redshift 1.0 to 0.5.
  This is the first time 
  a growth in the underlying halo mass has been measured {\it at high redshifts} within a single data
  survey, and provides evidence for the rapid accretion phase of massive halos.
  The mass accretion history follows the 
  form given in Wechsler et al. (2002) with $M(z) = M_0 e^{-\beta z}$, where
  $\beta \sim 1.07 \pm 0.57 (1.54 \pm 0.13)$ when only the VVDS points were used and 
  $\beta \sim 1.94 \pm 0.10 (2.09 \pm 0.04)$ 
  after including the SDSS data as reference points at low redshift
  and depending on the model used to obtain the best fits. 
  The addition of the low redshift SDSS points adds complications due to the 
  addition of possible systematics by comparing data from two different rest-frame bands,
  even after conversion to a common fiducial band. We adopt the average value of 
  $\beta \sim 1.3 \pm 0.30$
  from the VVDS points when discussing a growth in halo mass, and found to 
  be slightly higher than the results from N-body simulations.
  %further constrains the value for $\beta$ and diminishes the error bars.

  If we express this result in terms of the expected halo mass 
  at present times, $M_0 \simeq$
  $10^{13.5} h^{-1}M_\odot$, such halos appear to accrete $m \sim$ 0.25 $M_0$
  between redshifts of 0.5 and 1.0. Stewart et al. (2007) have shown that $\sim$
  25\% (80\%) of $M_0=10^{13} h^{-1}M_\odot$ halos experienced an $m > 0.3 M_0$ 
  ($m > 0.1 M_0$) merger event in the last 10 Gyr,
  this would translate into a $m > 0.1 M_0$ merger event
  over the redshift range z=[0.5-1.0] for the high mass halos here. 
  From merger rate studies one finds that 30\% of the stellar mass 
  of massive galaxies with $10^{10} M_\odot < M < 10^{11} M_\odot$ has been
  assembled through mergers since z=1 (de Ravel et al. 2009, 
  and references therein). The integrated stellar mass growth obtained can
  then be compared to the halo mass growth obtained here.

  For samples at similar redshifts we see that the average halo mass, $<M>$, 
  generally increases with
  the luminosity threshold of the sample, with a very mild hint
  of a decreasing galaxy satellite fraction. 
  This implies that galaxies in the faint sample show a stronger probability of being
  satellites in low mass halos as compared to bright galaxies in massive halos.

  We also find that the satellite fraction or average number of satellite galaxies
  appears to slowly increase over the redshift interval [0.5,1.0], but a stronger
  increase by a factor of $\sim$3 over z=[0.1,0.5] is seen. 
  %This increase in the number of satellite galaxies can be explained by 
  %the dynamical friction of subhalos within their parent halos (Conroy et al. 2006).
  This can be understood in terms of the dynamical friction that subhalos hosting
  satellite galaxies encounter
  within their host halos. The efficiency of dynamical friction 
  depends on the relative subhalo to halo mass. 
  Subhalos experience more efficient dynamical friction in low
  mass halos, which can be thought of as progenitor halos at high redshift. 
  %These halos at high redshift would have a tendency to host fewer satellite 
  %galaxies contained in subhalos. 
  The subhalos are continuously subjected to tidal stripping
  and gravitational heating within the dense environments and get eroded if not completely.
  As time evolves the halo accretes mass and undergoes mergers with other halos. 
  The subhalos that form
  as remnants of halo mergers are now more likely to remain intact within the higher
  mass halo, in turn leading to a larger number of satellite galaxies in
  present-day halos.

  A comparison with the SDSS results shows a few interesting features.  
  The value for $M_{min}$, which is the mass of a halo hosting at least one
  central galaxy on average, in $40 \%$ of the luminosity threshold VVDS samples 
  is similar to values for local SDSS galaxies. 
  Whereas, $M_1$ is generally higher for  VVDS galaxies as
  compared to what is seen locally. 
  %This is probably linked to previous
  %results demonstrating a correlation function for similar faint galaxy
  %samples that is lower than local SDSS and 2dF galaxies (Pollo et al. 2006).
  The ratio of $M_1/M_{min}$ is found to be considerably
  higher (almost a factor of 2) in the VVDS as compared to the SDSS results.
  This shows that in order to begin hosting satellite galaxies, halos at high redshift need to
  accrete a larger amount of mass than is seen locally. Hence one would
  observe roughly
  twice as many local satellite galaxies than high redshift ones within the
  same evolved halo mass. This is another line of evidence in favor of the lower 
  observed satellite fraction at high redshift and high local satellite fraction.
  This interpretation is
  highly simplified in light of the fact that the results have been  
  obtained with data taken in different restframe bands. 

  In order to investigate further and better constrain the mass growth and
  evolution in the number of satellite galaxies per halo over a larger redshift
  range, one needs to have
  samples from the same survey at low redshifts. This can be done with samples
  from deeper and wider redshift surveys. 
  Here we have concentrated on luminosity-threshold samples leading
  to a link between the luminosity of galaxies and the underlying dark matter
  distribution. The present paper can be seen as a precursor to many studies that can 
  be carried out with larger samples than the VVDS, 
  including CLF - conditional luminosity function studies (e.g. van den Bosch et al. 2003, etc.),
  analyses with galaxy samples of different stellar masses (Zheng et al. 2007), etc..
  They will certainly add
  to the understanding of the vast pool of underlying dark matter properties and hopefully obtain
  tighter constraints on models of galaxy formation.

\section{Acknowledgements}
  UA would like to acknowledge funding from the Marie Curie training network supported 
  by the European Community's Sixth Framework Programme (FP6). 
  UA also thanks Alessandro Sozzetti and Martin Kilbinger for helpful discussions.
  The authors thank the anonymous referee for very useful comments and suggestions
  that helped improve the paper.
  This research program has been developed within the framework of the VVDS
  consortium. 
  This work has been funded in part by the ANR program ANR-05-BLAN-0283
  and partially supported by the CNRS-INSU and its
  Programme National de Cosmologie (France), and by the Italian Ministry (MIUR)
  grants COFIN2000 (MM02037133) and COFIN2003 (No. 2003020150) and by INAF
  grants (PRIN-INAF 2005) and the grant of Polish Ministry of Science and
  Higher Education PBZ/MNiSW/07/2006/34.
  The VLT-VIRMOS observations have been carried out on guaranteed time (GTO)
  allocated by the European Southern Observatory (ESO) to the VIRMOS consortium,
  under a contractual agreement between the Centre National de la Recherche
  Scientifique of France, heading a consortium of French and Italian institutes,
  and ESO, to design, manufacture and test the VIMOS instrument.

\section*{}

  {\large \bf List of Affiliations} \\
      $^1$Laboratoire d'Astrophysique de Marseille, UMR 6110 CNRS-Universit\'e de Provence, BP8, F-13376 Marseille Cedex 12, France \\%1 
      $^2$INAF-Osservatorio Astronomico di Brera, Via Brera 28, I-20021, Milan, Italy\\ %2
      $^3$Max Planck Institut f\"ur Extraterrestrische Physik (MPE), Giessenbachstrasse 1,
     D-85748 Garching bei M\"unchen,Germany\\ %3 
      $^4$Max Planck Institut f\"ur Astrophysik, D-85741, Garching, Germany\\ %4 
      $^5$Centre de Physique Th\'eorique, UMR 6207 CNRS-Universit\'e de Provence, F-13288, Marseille, France\\ %5
      $^6$Universit\"atssternwarte M\"unchen, Scheinerstrasse 1, D-81679 M\"unchen, Germany\\ %6
      $^7$The Andrzej Soltan Institute for Nuclear Studies, ul. Hoza 69, 00-681 Warszawa, Poland\\ %8
      $^8$Astronomical Observatory of the Jagiellonian University, ul Orla 171, PL-30-244, Krak{\'o}w, Poland\\ %6 now 7
      $^9$INAF-Osservatorio Astronomico di Bologna, Via Ranzani 1, I-40127, Bologna, Italy\\ %7 now 9 
      $^{10}$IASF-INAF, Via Bassini 15, I-20133, Milano, Italy\\ %8 now 10
      $^{11}$IRA-INAF, Via Gobetti 101, I-40129, Bologna, Italy\\ %9 now 11
      $^{12}$INAF-Osservatorio Astronomico di Capodimonte, Via Moiariello 16, I-80131, Napoli, Italy\\ %10 now 12
      $^{13}$Universit\`a di Bologna, Dipartimento di Astronomia, Via Ranzani 1, I-40127, Bologna, Italy\\ %11 now 13
      $^{14}$Laboratoire d'Astrophysique de Toulouse/Tabres (UMR5572), CNRS, Universit\'e Paul Sabatier - Toulouse III, \\
             Observatoire Midi-Pyr\'en\'ees, 14 av. E. Belin, F-31400, Toulouse, France\\  %12 now 14
      $^{15}$Institut d'Astrophysique de Paris, UMR 7095, 98 bis Bvd Arago, F-75014, Paris, France\\  %13 now 15
      $^{16}$Observatoire de Paris, LERMA, 61 Avenue de l'Observatoire, F-75014, Paris, France\\ %14 now 16
      $^{17}$Astrophysical Institute Potsdam, An der Sternwarte 16, D-14482, Potsdam, Germany\\ %15 now 17
      $^{18}$INAF-Osservatorio Astronomico di Roma, Via di Frascati 33, I-00040, Monte Porzio Catone, Italy\\ %16 now 18
      $^{19}$Universit\'a di Milano-Bicocca, Dipartimento di Fisica, Piazza delle Scienze 3, I-20126, Milano, Italy\\  %17 now 19
      $^{20}$Integral Science Data Centre, ch. d'\'Ecogia 16, CH-1290, Versoix, Switzerland\\ %18 now 20
      $^{21}$Geneva Observatory, ch. des Maillettes 51, CH-1290, Sauverny, Switzerland\\ %19 now 21
      $^{22}$Centro de Astrof{\'{i}}sica da Universidade do Porto, Rua das Estrelas, P-4150-762, Porto, Portugal\\ %20 now 22
      $^{23}$Institute for Astronomy, 2680 Woodlawn Dr., University of Hawaii, Honolulu, Hawaii, 96822, USA\\ %21 now 23
      $^{24}$School of Physics \& Astronomy, University of Nottingham
	    University Park, Nottingham, NG72RD, UK\\ %22 now 24
      $^{25}$Canada France Hawaii Telescope corporation, Mamalahoa Hwy,  
     Kamuela, HI-96743, USA\\ %23 now 25
      $^{26}$INAF - Osservatorio Astronomico di Torino, Strada Osservatorio 20, 
      Pino Torinese, I-10025, Italy % 26

\section{Appendix}

    Here we present the table of values for the projected correlation function ($w_p$)
    and associated errors ($\sigma$) at different values of $r_p$ (in units of Mpc $h^{-1}$) 
    for the different subsamples mentioned in Table~\ref{table1} in the main text. 
    The values for $w_p$ and $\sigma$ are reported horizontally at the corresponding values
    for $r_p$ in the top row.
   \begin{table*}
     \caption{Projected correlation function with associated errors 
	at different $r_p$ for the different subsamples}
%     \caption{Different subsamples for selection 1}             % title of Table
     \label{table_a}      % is used to refer this table in the text
     \centering                          % used for centering table
     \begin{tabular}{l l c c c c c c c c c c c}      
       \hline\hline                 
       $M_B^{low}$ & $M_B^{high}$ & 0.13 & 0.25 & 0.40 & 0.63 & 1.00 & 1.58 & 2.51 & 3.98 & 6.31 & 10.00 & $r_p$ \\
       \hline  % inserts single horizontal line
        $-17.31$ & $-17.77$ & 111.66 & 82.28  & 47.85 & 36.06 & 27.95 & 15.88 & 14.05 & 10.51 & 5.96 & 5.84 & $w_p$ \\
                 &          &  37.60 & 29.03  & 16.56 & 10.86 & 9.12  & 6.86 & 5.05 & 3.48 & 2.51 & 2.29 & $\sigma$ \\ 
        $-17.66$ & $-18.00$ & 126.04 & 64.42 & 39.47 & 42.53 & 29.04 & 23.08 & 14.54 & 9.81 & 6.34 & 5.50 & $w_p$\\
		 &          &  30.37 & 19.46 & 9.63 & 7.13 & 5.98 & 4.16 & 2.84 & 2.19 & 1.70 & 1.20 & $\sigma$ \\ 
       \hline
        $-17.69$ & $-18.27$ & 144.25 & 92.38 & 53.99 & 50.06 & 38.30 & 24.86 & 19.11 & 12.79 & 8.34 & 7.75 & $w_p$\\
                 &          &  38.60 & 27.52 & 14.16 & 9.26 & 9.81 & 6.69 & 3.95 & 3.11 & 2.94 & 1.80 & $\sigma$ \\    
        $-18.16$ & $-18.50$ & 117.06 & 74.55 & 50.91 & 42.15 & 29.02 & 22.90 & 16.15 & 12.23 & 8.27 & 6.27 & $w_p$ \\
                 &          &  30.44 & 21.94 & 13.42 & 6.66 & 6.67 & 4.67 & 3.38 & 2.72 & 2.11 & 1.08 & $\sigma$ \\
       \hline
        $-18.02$ & $-18.71$ & 128.23 & 88.16 & 49.62 & 53.91 & 33.02 & 27.19 & 18.93 & 12.82 & 8.79 & 7.89 & $w_p$ \\
                 &          &  44.93 & 25.76 & 14.40 & 10.92 & 8.82 & 5.67 & 3.95 & 2.71 & 2.65 & 1.41 & $\sigma$ \\
        $-18.60$ & $-19.00$ & 102.02 & 62.36 & 54.98 & 29.06 & 23.63 & 21.53 & 14.76 & 11.98 & 8.09 & 6.29 & $w_p$ \\
                 &          &  32.39 & 22.12 & 13.33 & 8.43 & 8.13 & 5.01 & 3.70 & 3.07 & 2.37 & 1.23 & $\sigma$ \\
       \hline
        $-18.35$ & $-19.16$ & 101.78 & 99.94 & 72.26 & 45.95 & 32.67 & 25.17 & 19.58 & 14.98 & 10.30 & 9.06 & $w_p$\\
                 &          &  39.40 & 27.35 & 20.75 & 9.48 & 9.99 & 5.75 & 4.33 & 3.53 & 2.56 & 1.57 & $\sigma$ \\
        $-19.04$ & $-19.50$ & 129.74 & 95.63 & 68.13 & 30.44 & 27.86 & 21.47 & 16.47 & 13.56 & 10.41 & 6.53 & $w_p$ \\
                 &          &  44.85 & 30.38 & 17.90 & 8.77 & 8.93 & 6.04 & 3.99 & 3.20 & 2.39 & 1.28 & $\sigma$ \\
       \hline
        $-18.67$ & $-19.48$ & 98.56 & 84.77 & 84.00 & 55.49 & 36.81 & 24.64 & 21.98 & 15.27 & 10.22 & 8.18 & $w_p$ \\
                 &          & 46.76 & 40.82 & 25.60 & 15.59 & 12.05 & 7.89 & 5.94 & 4.37 & 3.25 & 1.95 & $\sigma$ \\
        $-19.37$ & $-20.00$ & 190.62 & 118.71 & 90.48 & 44.62 & 37.33 & 28.48 & 21.63 & 16.75 & 12.71 & 8.13 & $w_p$ \\
                 &          &  73.55 & 42.16 & 21.57 & 12.33 & 13.63 & 7.79 & 5.60 & 3.70 & 2.94 & 1.61 & $\sigma$ \\
       \hline\hline
     \end{tabular}
   \end{table*}

\end{document}